# Statistical model concept to quantify input and output of water, nitrogen and phosphorus for lakes with partly gauged watersheds

***Peter Borgen Sørensen*[*1], Anders Nielsen[2]***

[1]*Department of Ecoscience, Faculty of Technical Sciences, Aarhus University, C.F. Møllers Allé, Building 1120, 8000 Aarhus, Denmark.*

[2]*WateriTech, Døjsøvej 1, 8660 Skanderborg, Denmark*

[*]*pbs@ecos.au.dk*

## Abstract

Valid mass load predictions of nutrients, in particular nitrogen (N) and phosphorus (P), are needed for the limnological understanding of single lake ecosystems as well as larger river/lake ecosystems. The mass of N and P that enters a lake will determine the ecological state of the lake, and the mass release from the lake will determine the ecological state of downstream ecosystems. Hence, establishing sound quantifications of the external load is crucial and e.g. contributes to the foundation of assessments of necessary management interventions to improve or preserve the ecological integrity of lakes. The external load of N and P is an integral of several pathways, each having different contributions to the total mass load. Around the world, balances of N and P have been derived for decades to support both lake water quality monitoring and research, but it can be difficult and, thus, costly to make detailed and sufficiently covering measurement campaigns in all tributaries (surface as well as groundwater) in the watershed of the N and P load including seasonality and temporal change from year to year. Thus, load prediction is facing challenge of uncertainty due to unmeasured loads, which can be a consequence of limited resources available for the water flow recordings and water concentration measurements in inlets around the lake, or simply due to invisible water flow taking place through the lake bottom. The lake outlet will typically take place in one single river, so the outlet recording seems easier to measure than inlets, however, the outlet may also have unmeasured parts in cases where water is leaching out though the lake bottom. In this paper, we propose a method that applies incomplete data sets (incomplete in the sense of temporal frequency and percentage of gauged watershed) to generate time series that predict the N and P loads entering and leaving the lake.

## Introduction

The estimation of the nutrient load to lakes of nitrogen (N) and phosphorus (P) is a fundamental task needed for developing ecological models of lake processes. The uncertainty of the load estimates directly influences the uncertainty of the lake models, and therefore the predictions of the load need to be refined as much as possible in order to ensure valid lake modeling.

Specifically, the purpose of this method is to deliver optimal use of existing data to quantify N and P concentrations of the inflowing water as well as the water volumes entering and leaving the lake to predict the load of N and P. The major challenge is how to assess the unmeasured fraction of water, N and P.

The method has been developed as a part of the Danish National Monitoring Program (NOVANA), Svendsen et al., 2005a and 2005b and the data in the case study was obtained from this program.

## Methods

The principle is based on two linear model assumptions: (1) A mass balance of water, where the unmeasured fraction of water is predicted and separated into a baseflow and a surface-near flow, respectively; (2) A mass balance of N and P, where the measured concentration level is decomposed into a baseflow water concentration, a surface-near water concentration and a contribution from point sources. The predictions from these two linear models are used together with point source estimate, to predict the unmeasured load of N and P into the lake and downstream from the lake.

## Water model

A mass balance of water during one month for the lake is as follows:

$$\Delta V_{i,j} = Qin_{mea}|_{i,j} + Qin_{unmea}|_{i,j} + A \cdot \left(q_{pre}\big|_{i,j} - q_{eva}|_{i,j}\right) - Qout_{mea}|_{i,j} - Qout_{unmea}|_{i,j} \qquad (1)$$

Where

*i,j*: indexes of respectively year and month

$\Delta V_{i,j}$: Change in water volume in the lake for one month ($m^3$/month)

$Qin_{mea}|_{i,j}$: Measured amount of inflow water for one month ($m^3$/month)

$Qin_{unmea}|_{i,j}$: Unmeasured amount of inflow water for one month ($m^3$/month)

$q_{pre}\big|_{i,j}$: Precipitation for one month (m/month)

$q_{eva}|_{i,j}$: Evaporation for one month (m/month)

*A*: Lake area ($m^2$)

$Qout_{mea}|_{i,j}$: Measured water outflow in the outlet of the lake for one month ($m^3$/month)

$Qout_{unmea}|_{i,j}$: Unmeasured water outflow from the lake (leaching from bottom) for one month ($m^3$/month)

Eq. 1 can be rearranged as:

$$Qout_{mea}|_{i,j} = A \cdot \left(q_{pre}\big|_{i,j} - q_{eva}|_{i,j}\right) + Qin_{mea}\Big|_{i,j} + Qin_{unmea}|_{i,j} - Qout_{unmea}|_{i,j} - \Delta V_{i,j} \qquad (2)$$

The $Qout_{mea}|_{i,j}$ can be measured using a flow recorder at the outlet of the lake and the precipitation can be measured at a nearby station or estimated using methodological data. Evaporation can be measured using the energy balance of incoming energy balance. The $Qout_{unmea}|_{i,j}$ is the unmeasured outflow that can take place as net leaching from the lake bottom. The volume of water exchange due to precipitation and evaporation is estimated as

$$Qin_{vertical}|_{i,j} = A\left(q_{pre}\big|_{i,j} - q_{eva}|_{i,j}\right) \qquad (3)$$

Where $Qin_{vertical}|_{i,j}$ is the net volume of water supply into the lake for one month due to precipitation and evaporation. The $q_{pre}\big|_{i,j}$ variable is estimated using meteorological data, and $q_{eva}|_{i,j}$ is estimated using an energy balance by using methodological data in from of the energy radiation.

$$Qout_{mea}|_{i,j} = Qin_{vertical}|_{i,j} + Qin_{mea}|_{i,j} + Qin_{unmea}|_{i,j} - Qout_{unmea}|_{i,j} - \Delta V_{i,j} \qquad (4)$$

There are two challenges displayed in Eq. 3. The volume change of water in the lake is rarely measured in Denmark and, thus, in praxis is unknown, the following linear reservoir model for the right-hand side of Eq. 4 is defined:

$$Qout_{mea}|_{m} = \sum_0^l [F_t \cdot (Qin_{mea}|_{m-t} + Qin_{unmea}|_{m-t} + Qin_{vertical}|_{m-t})] - Qout_{unmea}|_{m} \qquad (5)$$

Where t is an index that counts time steps as months, and *m* is an index counting time steps from i= and j=1 as

$$m = 12 \cdot (i-1) + j \qquad (6)$$

and the linear reservoir weighting factor $F_t$ is defined as the fraction of the inflow at month *m* that sequentially will contribute to the outflow the same month (*t*=0), the next month (*t*=1), the month after next month (*t*=2), etc. The index $l$ is the number of months that are considered to have additional outflow because of an inflow at month *m*. Thus, the $F_l$ factor will sum up to

$$\sum_0^l F_z = 1 \qquad (7)$$

The linear reservoir model is derived based on a box model having continued input of a mass unit during a time unit, and where the output from the box is proportional to what is still left in the box, see Figure 1.

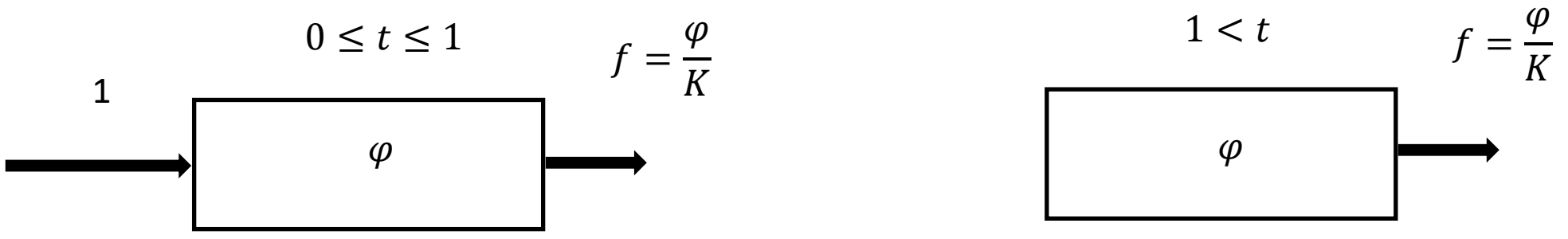

Figure 1. A linear reservoir model. To the left, the mass unit is introduced into the box continuously during a time unit. On the right, the condition for t>1 is shown, where the mass is released from the box proportionally to the remaining mass. Parameter *K* is a reservoir constant that controls the release rate and, thus, the time scale of the response for the reservoir.

$$\frac{d\varphi}{dt} = 1 - \frac{\varphi}{K} \;, \quad \text{for } 0 \le t \le 1 \qquad (7)$$

$$\frac{d\varphi}{dt} = -\frac{\varphi}{K} \;, \quad \text{for } 1 < t \qquad (8)$$

The initial condition for Eq 6 is $\varphi = 0$ for *t*=0, and the solution of Eq. 6, 7 is:

$$\varphi = K \cdot \left(1 - e^{-\frac{t}{K}}\right), for\ 0 \le t \le 1 \qquad (9)$$

$$\varphi = \varphi_{t=1} \cdot e^{-\frac{t-1}{K}}, for\ 1 < t \qquad (10)$$

The flux of mass out of the box (*f*) can now be calculated using the relationship $f = \frac{\varphi}{K}$ as

$$f = \left(1 - e^{-\frac{t}{K}}\right), for\ 0 \le t \le 1 \qquad (11)$$

$$f = \frac{\varphi_{t=1}}{K} \cdot e^{-\frac{t-1}{K}}, for\ 1 < t \qquad (12)$$

The mass leaving the box during the initial time unit ($0 \le t \le 1$) can be calculated using Eq 11 as

$$F_0 = \int_0^1 f dt = 1 + K \cdot \left(e^{-\frac{1}{K}} - 1\right) \qquad (13)$$

The mass that leaves the box after the initial time unit during the period from *t* to *t*+1 can be derived using Eq 12 and 11 for t>1 as

$$F_t = \int_t^{t+1} f d\omega = K \cdot \left(1 - e^{-\frac{1}{K}}\right)\left(e^{-\frac{t-1}{K}} - e^{-\frac{t}{K}}\right) \qquad (14)$$

Where $F_t$ is the fraction of the inflow at step *m-t* that is contributing to the outlet at step *m*. The principle is shown in Figure 2 for *l*=10 and two *K* values (0.5 and 2). Thus, if a lake is estimated to have *K*=0.5, then the major part of inflowing water will also result in an increased outflow in the same month, however, a fraction of 0.38 will lead to an increase the month after, and a small fraction will lead to increased outflow two months after. For *K*=2, the delay of the of the response to an input is strong, and Figure 2 shows that the increase in I outflow is largest the month following the month when inflow took place. Thus, the *K* parameter characterizes the hydraulic property of the lake, where a lake having a large inflow of water and a small surface area will tend to have small values of *K*, while larger lakes having larger surface limited inflow will tend to have larger *K* values. The physical property of the outlet will also play a role, where a wide outlet will tend to keep the lake volume constant and therefore result in a small value of *K* in contrast to a narrower outlet.

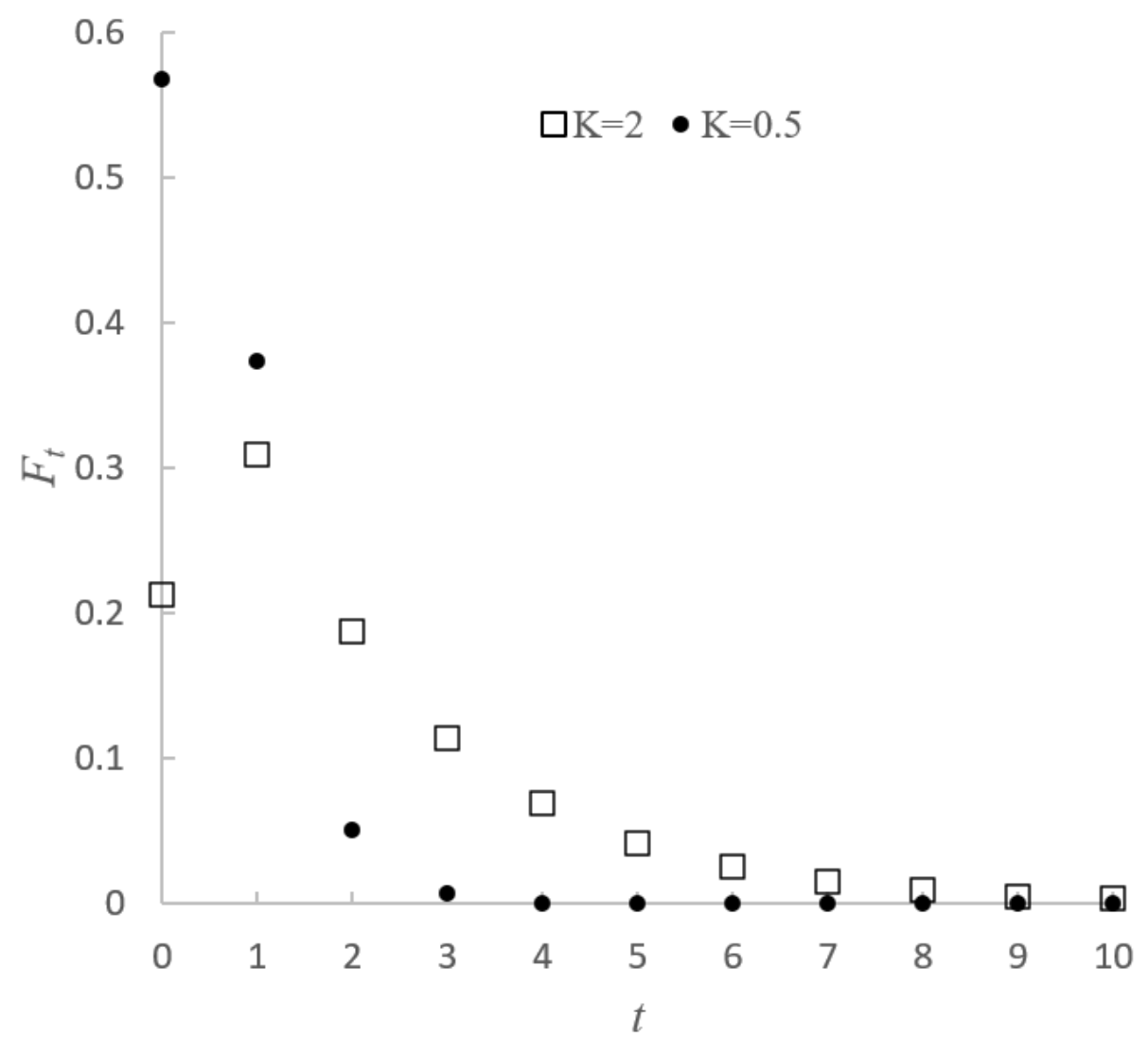

Figure 2. Numerical example for *l*=10 and the *K* values 0.5 and 2.

The unmeasured part of the water volume needs to be estimated in Eq. 5, and for this purpose the following linear model is used:

$$Qin_{unmea}|_m = \alpha_n \cdot Qin_{mea}|_m + \beta 1_i \cdot \gamma_j \tag{15}$$

Where $\alpha_n$ is the slope between the measured and unmeasured flow for the $n^{th}$ combination of measuring station location. In some cases, the locations of the water flow stations are changed and a slope needs to be estimated for each new location in order to take into account that this may influence the fraction of unmeasured water. The parameter $\beta 1_i$ estimates a yearly water volume entering the lake during the year *i* without having any relation to $Qin_{mea}|_m$. This could be ground water surplus to the lake through the bottom or from aquifers adjacent to the lake perimeter. The parameter $\gamma_j$ describes the seasonal distribution of this surplus water volume due to seasonality of the ground water table. Thus, the following constrain applies to $\gamma_j$:

$$\sum_{j=1}^{12} \gamma_j = 1 \tag{16}$$

The unmeasured outflow of water is assumed to be seepage through the lake bottom, as lakes rarely have more than one outlet at the surface. This seepage is assumed to be constant during the year, but may differ between years. The argument is that the seepage will tend to be rather constant due to the relatively fixed water table in the lake, however, during a longer period of several years it may change. So, the unmeasured outflow is estimated as a constant each year:

$$Qout_{unmea}|_m \approx \beta 2_i \ , \ \beta 2_i \geq 0 \tag{17}$$

The sign $\beta 2_i$ is restricted to conditions where unmeasured water is leaving the lake. Eqs. 5, 15 and 17 are combined in the statistical model for the measured outflow as:

$$Qout_{mea}|_m \approx \sum_0^l \left[F_t \cdot \left((1+\alpha_n) \cdot Qin_{mea}|_m + \beta 1_i \cdot \gamma_j\right)\right] - \beta 2_i \ , \ \beta 1_i, \beta 2_i \geq 0 \wedge [\beta 1_i > 0 \rightarrow \beta 2_i = 0] \tag{18}$$

Where the linear reservoir factor $F_t$ is estimated using Eq 13 and 14. Eq. 18 is used in a Maximum Likelihood estimate of the parameters. However, for practical fitting there is a challenge using Eq. 18 without including the constraints for $\beta 1_i$ and $\beta 2_i$. This is because the value of respectively $\beta 1_i$ and $\beta 2_i$ can compensate for each other such that an increase in the $\beta 1_i$ value can be compensated by a similar decrease in the $\beta 2_i$ value. The interpretation of this phenomenon is that if constant unmeasured water flow into and out of the lake exist simultaneously, then this is hidden for the mass balance governing the model.

The model assumes that the month having the least measured inflow of water is the month where the contribution from the surface-near water flow is negligible as an inflow contributor, this month is denoted the reference month for the specific year. Thus, every year has a reference month that may be different from year to year, denoted $jref_i$. The measured baseflow as a fraction of the inflow for each month is estimated based on the reference month:

$$Qin_{meabase}|_{i,j} \approx Qin_{mea}|_{jref_i} \tag{19}$$

The measured surface-near water flow is now predicted as:

$$Qin_{measurface}\big|_{i,j} \approx Qin_{mea}|_{i,j} - Qin_{meabase}|_{i,j} \tag{20}$$

The unmeasured base-flow entering the lake is predicted as:

$$Qin_{unmeabase}|_{j,i} \approx \alpha_n * Qin_{meabase}|_{i,j} + \beta 1_i \cdot \gamma_j \tag{21}$$

In Eq. 21, it is assumed that the volume of water entering the lake independently of the measured inlet volume is a baseflow. The argument is that unmeasured surface-near inflow correlates close to the variation in measured inflow because the surface-near water tends to induce large variations in the inflowing water volume, being a short term respond to rainfall that often takes place in the entire large catchment and, thus, is effective also on unmeasured areas. The unmeasured surface-near flow entering the lake is thus predicted as:

$$Qin_{unmeasurface}\big|_{j,i} \approx \alpha_n * Qin_{measurface}\big|_{i,j} \tag{22}$$

The total baseflow entering the lake can now be predicted as

$$Qin_{base}|_{j,i} \approx Qin_{meabase}|_{i,j} + Qin_{unmeabase}|_{i,j} \tag{23}$$

The total flow of water from surface-near sources is predicted as

$$Qin_{surface}\big|_{j,i} \approx (1 + \alpha_n) * Qin_{measurface}\big|_{i,j} \tag{24}$$

Eq. 15 assumes that the fraction of baseflow in unmeasured inflow is similar to the fraction of base flow in the measured inflow. The total waterflow out of the lake is predicted as

$$Q_{out}|_{j,i} = Qout_{unmea}|_{i,j} + Qout_{mea}|_{i,j} \tag{25}$$

## N and P model

The load of N and P is divided in two categories: a measured part and an unmeasured part. The measured part is predicted as the measured concentration level in the measured water in the inflow, and the unmeasured part is predicted using a model as described in the following. A mass balance is established for the measured transport of N and P that divides the total measured load into three transport routes: (1) $Qin_{measurface}\big|_{i,j}$; (2) $Qin_{meabase}|_{i,j}$; (3) N and P entering from point sources in the measured part of the lake watershed ($P_{effmea}\big|_{i,j}$) as

$$Load_{i,j} \approx Cin_{tot,mea}\big|_{i,j} \cdot Qin_{mea}|_{i,j} + Cin_{unmeabase}|_{i,j} \cdot Qin_{unmeabase}|_{i,j} + Cin_{unmeasurf}\big|_{i,j} \cdot Qin_{unmeasurf}\big|_{i,j} + P_{unmea}|_{i,j} \tag{26}$$

The unmeasured concentrations are assumed equal to the measured concentration for the base flow and surface-near flow, respectively:

$$Cin_{unmeabase}|_{i,j} \approx Cin_{meabase}|_{i,j} \quad , \quad Cin_{unmeasurf}\big|_{i,j} \approx Cin_{measurf}\big|_{i,j} \tag{27}$$

Where $Cin_{meabase}|_{i,j}$ is the base-flow concentration in the measured inlets and $Cin_{measurf}\big|_{i,j}$ is the surface-near concentration in the measured inlet water. The measured concentration will obviously only be a total bulk concentration of the mixed inlet water between base-flow and surface-near water. Hence, the base-flow and surface-near flow components need to be estimated in the total inlet concentration of the mixed water.

The total mass balance provides:

$$Cin_{tot,mea}\big|_{i,j} \cdot Qin_{mea}|_{i,j} = Cin_{mea,base}\big|_{i,j} \cdot Qin_{mea,base}\big|_{i,j} + Cin_{mea,surf}\big|_{i,j} \cdot Qin_{mea,surf}\big|_{i,j} + P_{effmea}\big|_{i,j} \tag{28}$$

where $Cin_{tot,mea}\big|_{i,j}$ is the total bulk concentration measured in the inlet to the lake in the same water volume that was measured as $Qin_{mea}|_{i,j}$, the concentration in respectively base-flow and surface-near flow is $Cin_{mea,base}\big|_{i,j}$ and $Cin_{mea,surf}\big|_{i,j}$. The effective contribution from point sources in the inlet is added as $P_{effmea}\big|_{i,j}$ and defined as the contribution form point sources that have discharged to the measured inlet.

It is not possible to measure the concentration levels separately in the baseflow and in the surface-near flow, as the water is mixed up in the streams. However, these concentration levels can be predicted using parameters estimated by a statistical model as follows, where Eq 29 is rearranged to form:

$$Cin_{tot,mea}\big|_{i,j} = Cin_{mea,base}\big|_{i,j} \cdot \frac{Qin_{meabase}|_{i,j}}{Q_{inmea}|_{i,j}} + Cin_{measurf}\big|_{i,j} \cdot \frac{Qin_{measurf}|_{i,j}}{Qin_{mea}|_{i,j}} + \frac{P_{effmea}\big|_{i,j}}{Qin_{mea}|_{i,j}} \tag{29}$$

And this equation can be rewritten as

$$Cin_{totmea}|_{i,j} = Cin_{meabase}|_{i,j} \cdot \left(1 - X_{i,j}\right) + Cin_{measurf}\big|_{i,j} \cdot X_{i,j} + \frac{P_{effmea}\big|_{i,j}}{Qin_{mea}|_{i,j}} \quad where\ X_{i,j} = \frac{Q_{mea,surf}\big|_{i,j}}{Qin_{mea}|_{i,j}} \tag{30}$$

And further rewritten to the form:

$$Cin_{totmea}|_{i,j} = Cin_{meabase}|_{i,j} + \left(Cin_{measurf}\big|_{i,j} - Cin_{meabase}|_{i,j}\right) \cdot X_{i,j} + \frac{P_{effmea}\big|_{i,j}}{Qin_{mea}|_{i,j}} \tag{31}$$

The quantification of effective point sources' contribution to the load by measured inlet water will typically be uncertain, both due to the quantification of the point source strength, where the discharge takes place, and the retention from the position of the point sources' discharge to the point of measurement at the lake inlet. Thus, a parameter $\rho$ is defined as an effective point source factor as

$$P_{effmea}\big|_{i,j} = \rho \cdot P_{mea}|_{i,j} \tag{32}$$

Under the condition that the local point source strength is known the Eq. 15 has the following form, where the depended variable is:

$$Y_{i,j} = \theta o + \theta 1 \cdot t + \left(\gamma 1_i + \gamma 2_j\right) \cdot X_{i,j} + \rho \cdot \frac{P_{mea}|_{i,j}}{Qin_{mea}|_{i,j}} \tag{33}$$

Where $t$ is time since the first measurement, and $\rho$ is the measured impact form point sources. Comparing Eq. 31 and 33, it is seen that the measured baseflow concentration can be predicted as

$$Cin_{meabase}|_{i,j} \approx \theta o + \theta 1 \cdot t \tag{34}$$

And the measured surface-near fraction can be predicted as

$$Cin_{measurface}\big|_{i,j} \approx Cin_{meabase}|_{i,j} + \gamma 1_i + \gamma 2_j \tag{35}$$

Consequently, the base-flow concentration is assumed to change gradually with time as a linear and second order relation. The argument is that the base-flow concentration is related to ground water that typically takes several years to form and, thus, is attributed to a sluggish change during time. The surface-near concentration is more dynamic, both in relation to the season and from one year to another year.

As seen in Eq. 31, the point sources are included as part of the dependent variable Y. The contribution from these point sources is not directly measurable and must therefore be estimated as a contribution upstream in the catchment to a given measuring station.

The predicted concentration levels from Eq. 25 and 26 are latent in the way that they are not measurable quantities that can be validated. Thus, the predictions from Eq. 25 and 26 may be adjusted or replaced using expert judgement, if the values seem unrealistic and local knowledge about proper value intervals exists.

## Load prediction

The predicted water volumes and concentration levels are multiplied to predict the mass load in and out the lake:

$$Load_{in}|_{i,j} \approx Cin_{tot,mea}|_{i,j} \cdot Qin_{mea}|_{i,j} + Cin_{meabase}|_{i,j} \cdot Qin_{unmeabase}|_{j,i} + Cin_{measurf}|_{i,j} \cdot Qin_{unmeasurf}|_{j,i} + \rho \cdot P_{unmea}|_{i,j} + At|_{i,j} \quad (36)$$

Where $P_{unmea}|_{i,j}$ is the contribution from unmeasured point sources and $At|_{i,j}$ is the atmospheric deposition on the lake surface. The load out of the lake is estimated as

$$Load_{out}|_{i,j} \approx Cin_{meaout}|_{i,j} \cdot Q_{out}|_{j,i} \quad (37)$$

Where $Cin_{meaout}|_{i,j}$ is the measured concentration in the outlet.

# Results

## Case study Bryrup Langsø

The Danish Lake Bryrup Langsø was selected as an illustrative case. The lake is a small to middle sized shallow lake covering an area of 37 ha and with a mean depth of 4.6 m. The inlet measuring stations are located along the major inlet, but the station was relocated between 1995 and 1996, so the fraction of measured water may have changed as a consequence. Hence, the model concept introduced in this paper distinguishes between these two inlet stations (*n*=2). Three smaller ungagged streams enter the lake, as seen in Figure 1, so this indicates that the model should identify a fraction unmeasured water flow into the lake.

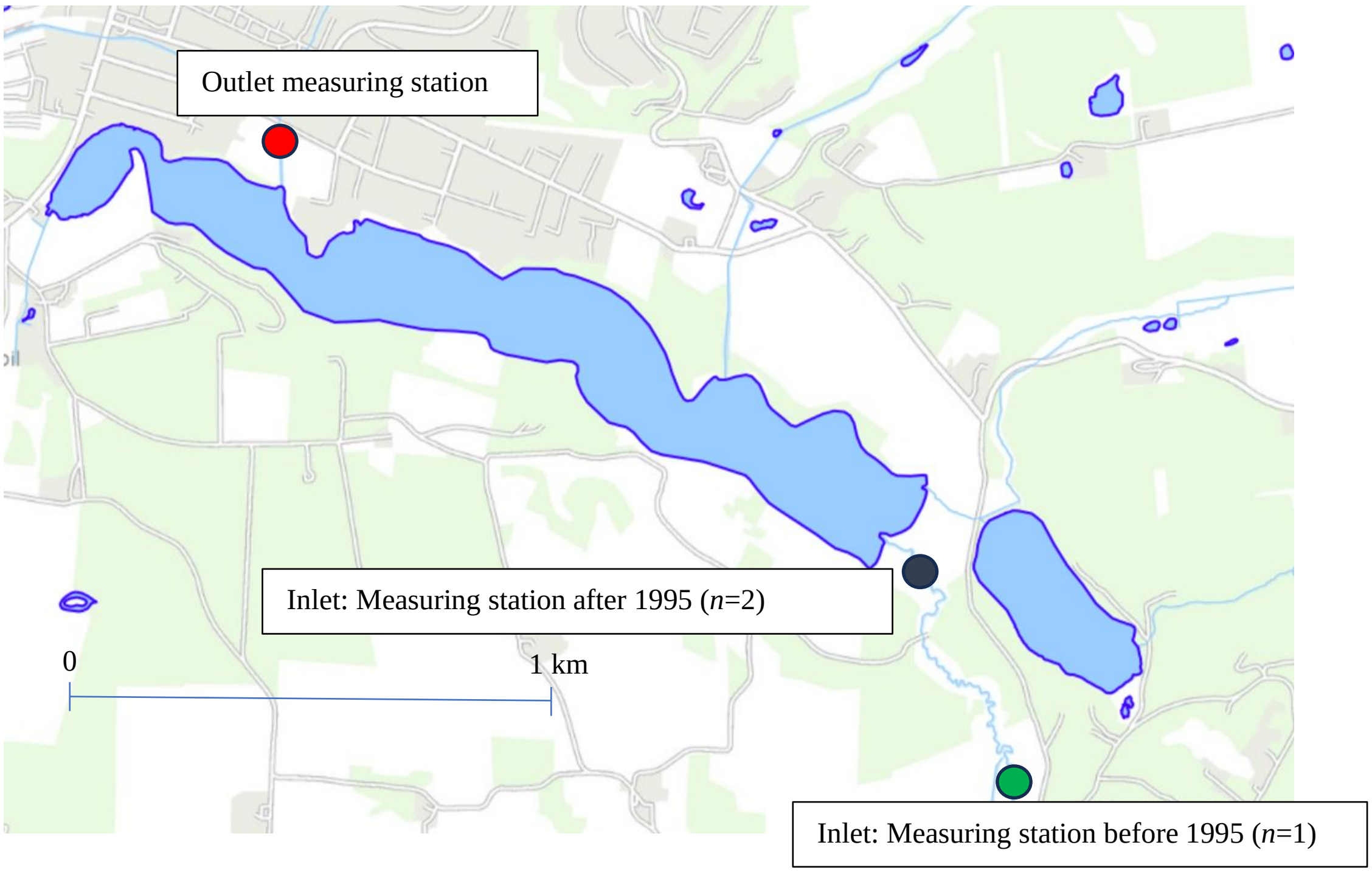

Figure 1. The overview of Bryrup Langsø case study. The lake is shaped as a rift formed by the ice age. The inlet stations are located at the main stream entering at the eastern end, and the outlet is located at the western end.

The data are collected at the stations as a part of the Danish National Monitoring Program (NOVANA), Svendsen et al., 2005a and 2005b, and the data in the case study are obtained from this program. The water flow is predicted using local calibrated correlations between water flow and depth. This results in continuous water flow prediction that is integrated to predict the water volume passing the cross section during one month. The total concentrations of N and P have been measured during the years, typically one or two bulk water samples per month. The measured concentration levels have been linear interpolated, and the mean value is multiplied with the paired water volume to predict the actual mass of transported matter for one month. Monthly data was collected in the period 1990-2020, as listed in Table A1, see appendix.

## Water model

The coefficients in the linear model of the water balance model (Eq. 4) are estimated in Table 1.

| coefficient | Estimate | unit | coefficient | Estimate | unit | coefficient | Estimate | unit |
|---|---|---|---|---|---|---|---|---|
| $\beta1_1$ | 900000 | $m^3$ | $\beta1_{28}$ | 1700000 | $m^3$ | $\beta1_{23}$ | 0 | $m^3$/month |
| $\beta1_2$ | 1040000 | $m^3$ | $\beta1_{29}$ | 1600000 | $m^3$ | $\beta1_{24}$ | 0 | $m^3$/month |
| $\beta1_3$ | 900000 | $m^3$ | $\beta1_{30}$ | 2120000 | $m^3$ | $\beta1_{25}$ | 0 | $m^3$/month |
| $\beta1_4$ | 860000 | $m^3$ | $\beta1_{31}$ | 2440000 | $m^3$ | $\beta1_{26}$ | 0 | $m^3$/month |
| $\beta1_5$ | 1660000 | $m^3$ | $\beta1_{32}$ | 1720000 | $m^3$ | $\beta1_{27}$ | 0 | $m^3$/month |
| $\beta1_6$ | 3660000 | $m^3$ | $\beta2_1$ | 0 | $m^3$/month | $\beta1_{28}$ | 0 | $m^3$/month |
| $\beta1_7$ | 920000 | $m^3$ | $\beta2_2$ | 0 | $m^3$/month | $\beta1_{29}$ | 0 | $m^3$/month |
| $\beta1_8$ | 560000 | $m^3$ | $\beta2_3$ | 0 | $m^3$/month | $\beta1_{30}$ | 0 | $m^3$/month |
| $\beta1_9$ | 1180000 | $m^3$ | $\beta2_4$ | 0 | $m^3$/month | $\beta1_{31}$ | 0 | $m^3$/month |
| $\beta1_{10}$ | 720000 | $m^3$ | $\beta2_5$ | 0 | $m^3$/month | $\beta1_{32}$ | 0 | $m^3$/month |
| $\beta1_{11}$ | 1620000 | $m^3$ | $\beta2_6$ | 0 | $m^3$/month | $\gamma_1$ | 0.073 | 1/month |
| $\beta1_{12}$ | 480000 | $m^3$ | $\beta2_7$ | 0 | $m^3$/month | $\gamma_2$ | 0.098 | 1/month |
| $\beta1_{13}$ | 1360000 | $m^3$ | $\beta2_8$ | 0 | $m^3$/month | $\gamma_3$ | 0.134 | 1/month |
| $\beta1_{14}$ | 1280000 | $m^3$ | $\beta2_9$ | 0 | $m^3$/month | $\gamma_4$ | 0.110 | 1/month |
| $\beta1_{15}$ | 1360000 | $m^3$ | $\beta2_{10}$ | 0 | $m^3$/month | $\gamma_5$ | 0.110 | 1/month |
| $\beta1_{16}$ | 1700000 | $m^3$ | $\beta2_{11}$ | 0 | $m^3$/month | $\gamma_6$ | 0.098 | 1/month |
| $\beta1_{17}$ | 980000 | $m^3$ | $\beta2_{12}$ | 0 | $m^3$/month | $\gamma_7$ | 0.073 | 1/month |
| $\beta1_{18}$ | 540000 | $m^3$ | $\beta2_{13}$ | 0 | $m^3$/month | $\gamma_8$ | 0.073 | 1/month |
| $\beta1_{19}$ | 920000 | $m^3$ | $\beta2_{14}$ | 0 | $m^3$/month | $\gamma_9$ | 0.061 | 1/month |
| $\beta1_{20}$ | 620000 | $m^3$ | $\beta2_{15}$ | 0 | $m^3$/month | $\gamma_{10}$ | 0.049 | 1/month |
| $\beta1_{21}$ | 100000 | $m^3$ | $\beta2_{16}$ | 0 | $m^3$/month | $\gamma_{11}$ | 0.061 | 1/month |
| $\beta1_{22}$ | 760000 | $m^3$ | $\beta2_{17}$ | 0 | $m^3$/month | $\gamma_{12}$ | 0.061 | 1/month |
| $\beta1_{23}$ | 1060000 | $m^3$ | $\beta2_{18}$ | 0 | $m^3$/month | $\alpha_1$ | 0.20 | - |
| $\beta1_{24}$ | 1280000 | $m^3$ | $\beta2_{19}$ | 0 | $m^3$/month | $\alpha_2$ | 0.21 | - |
| $\beta1_{25}$ | 880000 | $m^3$ | $\beta2_{20}$ | 0 | $m^3$/month | $K$: | 0.01 | - |
| $\beta1_{26}$ | 1360000 | $m^3$ | $\beta1_{21}$ | 0 | $m^3$/month | | | |
| $\beta1_{27}$ | 1800000 | $m^3$ | $\beta1_{22}$ | 0 | $m^3$/month | | | |

Table 1. Estimated values for the coefficients of the water balance model (Eq. 4) for the lake Bryrup Langsø. $R^2$=0.95.

The $\beta1_i$ values are total volume of water entering the lake during one year, independent of the variation of the measured inflow. The $\gamma_j$ values distribute $\beta1_i$ between the months. The water model interprets $\beta1_i$as a base flow, so the value should have correlation to the precipitation the year before, and this is investigated for Bryrup Langsø in Table 2, where $\beta1_i$ is used as a response variable for the linear combination of the yearly precipitation (mm/year), respectively the same year ($Prec_i$), the former year ($Prec_{i-1}$) and the year before the former year ($Prec_{i-2}$). If the intercept is related to the base-flow into the lake, then the intercept in January should be correlated to the precipitation the year up to January (year number $i$-1) and not the year that starts in January (year number $i$). This is seen to be the case in Table 2, where only the precipitation the year before the intercept is seen to be significant, and the year before the former year is not significant ($Prec_{i-2}$).

| Response variable: | $\beta1_i$ | | | |
|---|---|---|---|---|
| Prediction variables | Sum square | Df | F | Pr (>F) |
| $Prec_i$ | $91\cdot 10^6$ | 1 | 0.031 | 0.86 |
| $Prec_{i-1}$ | $17899\cdot 10^6$ | 1 | 6.147 | 0.02 |
| $Prec_{i-2}$ | $416\cdot 10^6$ | 1 | 0.143 | 0.71 |
| Residuals | $75714\cdot 10^6$ | 26 | | |

Table 2. ANOVA table testing the influence of different years' precipitation on the response variable $\beta_i$

The intercept variation between months ($\gamma_i$) is shown during the year in Figure 1. Bryrup Langsø is in a climatic zone having a seasonality, where the evaporation from the soil surface and vegetation is high during the summer period, such that the formation of surface-near water is limited in this part of the season. While the winter period is characterized by having much more precipitation than evaporation. Thus, the variation in Figure 1 is expected, where the baseflow to the lake is increasing during the winter period build up, having a top value in spring and followed by a decrease during the summer period.

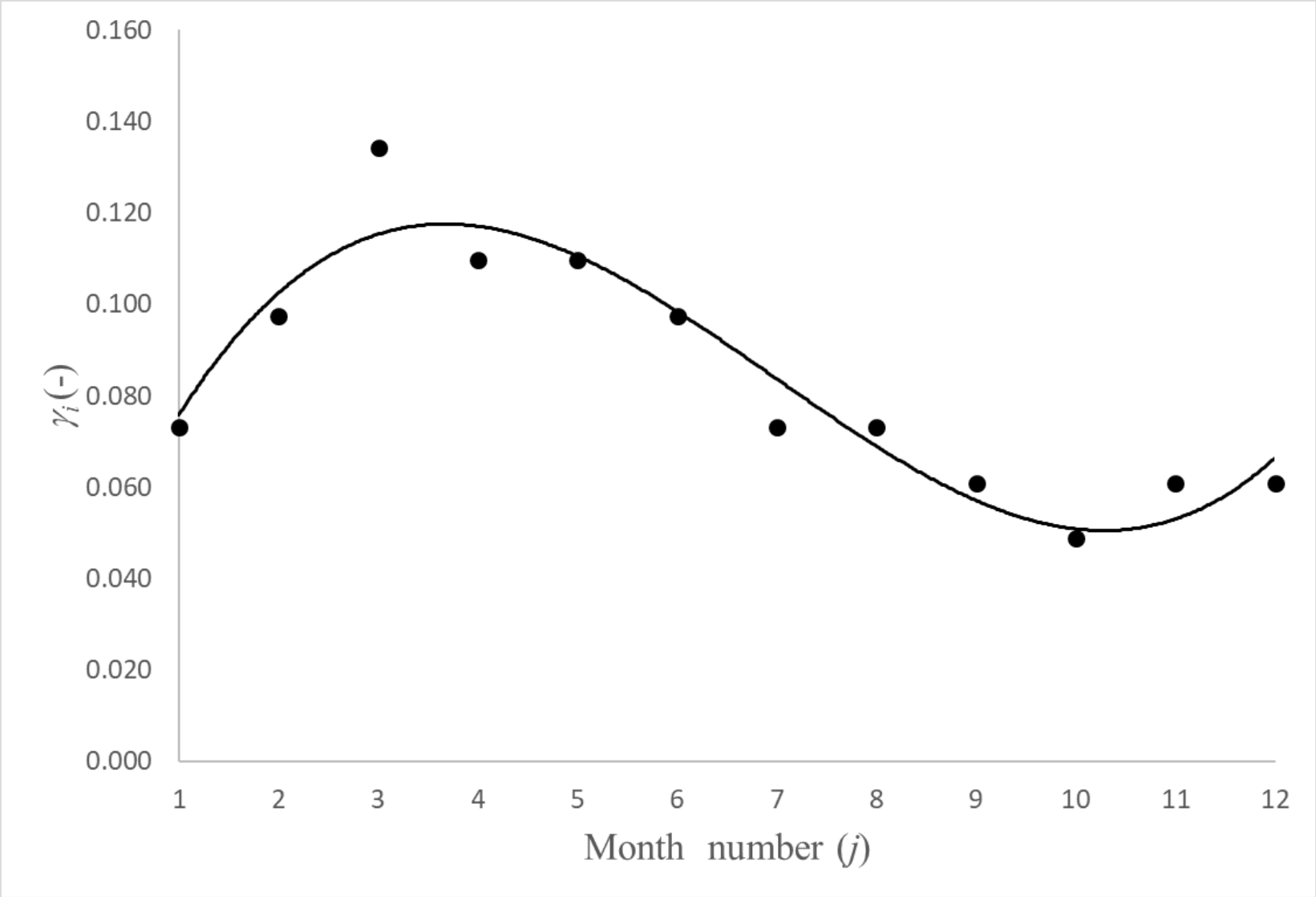

Figure 1. The variation in incept ($\gamma_i$) during the year.

The slopes ($\alpha$) in Table 1 are estimated for two different gauge locations corresponding to Table A1 in appendix A, where it is seen that station location "d" is changed to "e" from year 1995 to 1996. This is because the location of the inlet station was moved, so the fraction of respectively measured and unmeasured water in the inlet may also have changed. However, the slope value of $\alpha_1 = 0.20$ is nearly similar to the value of $\alpha_2 = 0.21$, so the change of station location does not indicate a considerable change in the fraction of unmeasured water. It is also seen that the value $K$=0.01 discloses a limited hydraulic memory effect from the former month, which, again, shows a high hydraulic load of water. The water volume inflow during one month divided by lake area is typically 1-3 m, indicating a high hydraulic load level on the lake and, thus, a highly limited hydraulically memory from the former month.

The general water volume budget is disclosed in Table 3. The unmeasured outflow ($Qout_{unmea}|_{i,j}$) is zero for the entire data set, so the lake does not have net release of water at any time, however, it is impossible to identify conditions where a simultaneous inflow and out flow exist through the lake bottom. The unmeasured base-flow can either come from wells outside the lake or form wells directly flowing into the lake from the lake bottom.

| Station location ($n$) | Mean monthly volume inflowing water (1000 $m^3$/month) | | | |
|---|---|---|---|---|
| | Reservoir type | Measured | Unmeasured | Total |
| 1 | Base flow | 200 | 129 | 330 |
| | Surface-near flow | 291 | 58 | 350 |
| | Total | 492 | 188 | 679 |
| 2 | Base flow | 214 | 152 | 366 |
| | Surface-near flow | 225 | 47 | 273 |
| | Total | 439 | 199 | 638 |

Table 3. The mean monthly inflow exclusive $Qin_{vertical}|_{i,j}$ is divided into respectively base-flow and surface-near flow and measured and unmeasured volumes. The inlet measuring station was relocated at the end of 1995, so $n$=1 covers the measurement before 1996 (1990-1995), and $n$=2 includes the measurements from year 1996 (1996-2021).

## N Model

The N concentration in the inlet is analyzed using equations 24-26. The coefficient table for Eq. 23 is shown in Table 3. The point source factor $\rho$ was estimated to be negative and non-significantly different from zero. Thus, the model could not detect any influence to the inlet due to point sources. In Denmark, the point sources for N are typically small compared to the leaching from soil, so a non-significant contribution from point sources is not surprising. This coincides with the comparison between the point source strength in Table A1 for measured catchment that has a level in the interval 50-100 kg/month, with the load from measured water in Figure 5, where the levels are much higher, at 1500-12000 kg/month.

| coefficient | Estimate | unit | coefficient | Estimate | unit | coefficient | Estimate | unit |
|---|---|---|---|---|---|---|---|---|
| $\gamma1_1$ | 0 | mg/l | $\gamma1_{17}$ | -1.51 | mg/l | $\gamma2_1$ | 1.24 | mg/l |
| $\gamma1_2$ | 0.15 | mg/l | $\gamma1_{18}$ | -2.05 | mg/l | $\gamma2_2$ | 1.00 | mg/l |
| $\gamma1_3$ | 3.21 | mg/l | $\gamma1_{19}$ | -2.25 | mg/l | $\gamma2_3$ | 0.29 | mg/l |
| $\gamma1_4$ | 0.95 | mg/l | $\gamma1_{20}$ | -0.88 | mg/l | $\gamma2_4$ | -0.12 | mg/l |
| $\gamma1_5$ | -0.59 | mg/l | $\gamma1_{21}$ | -1.21 | mg/l | $\gamma2_5$ | -0.39 | mg/l |
| $\gamma1_6$ | -0.66 | mg/l | $\gamma1_{22}$ | -2.44 | mg/l | $\gamma2_6$ | -1.47 | mg/l |
| $\gamma1_7$ | 2.79 | mg/l | $\gamma1_{23}$ | -2.17 | mg/l | $\gamma2_7$ | -1.54 | mg/l |
| $\gamma1_8$ | 1.82 | mg/l | $\gamma1_{24}$ | -1.81 | mg/l | $\gamma2_8$ | -1.56 | mg/l |
| $\gamma1_9$ | 2.33 | mg/l | $\gamma1_{25}$ | -2.68 | mg/l | $\gamma2_9$ | -0.91 | mg/l |
| $\gamma1_{10}$ | 0.07 | mg/l | $\gamma1_{26}$ | -2.02 | mg/l | $\gamma2_{10}$ | 0.05 | mg/l |
| $\gamma1_{11}$ | -0.81 | mg/l | $\gamma1_{27}$ | -2.12 | mg/l | $\gamma2_{11}$ | 1.37 | mg/l |
| $\gamma1_{12}$ | -0.52 | mg/l | $\gamma1_{28}$ | -1.26 | mg/l | $\gamma2_{12}$ | 1.36 | mg/l |
| $\gamma1_{13}$ | -1.30 | mg/l | $\gamma1_{29}$ | -0.92 | mg/l | $\theta o$ | 8.61 | mg/l |
| $\gamma1_{14}$ | 0.76 | mg/l | $\gamma1_{30}$ | 0.18 | mg/l | $\theta 1$ | -0.0050 | mg/(l·t) |
| $\gamma1_{15}$ | -0.13 | mg/l | $\gamma1_{31}$ | -2.02 | mg/l | $\rho$ | 0 | - |
| $\gamma1_{16}$ | -1.10 | mg/l | $\gamma1_{32}$ | 0.66 | mg/l | | | |

Table 4. Estimated parameters for Eq. (24), $R^2$=0.70.

The time series of predicted concentrations is shown in Figure 2 together with the measured values subtracted point source estimates. Some extreme values, especially some higher levels, seem poorly described by the model.

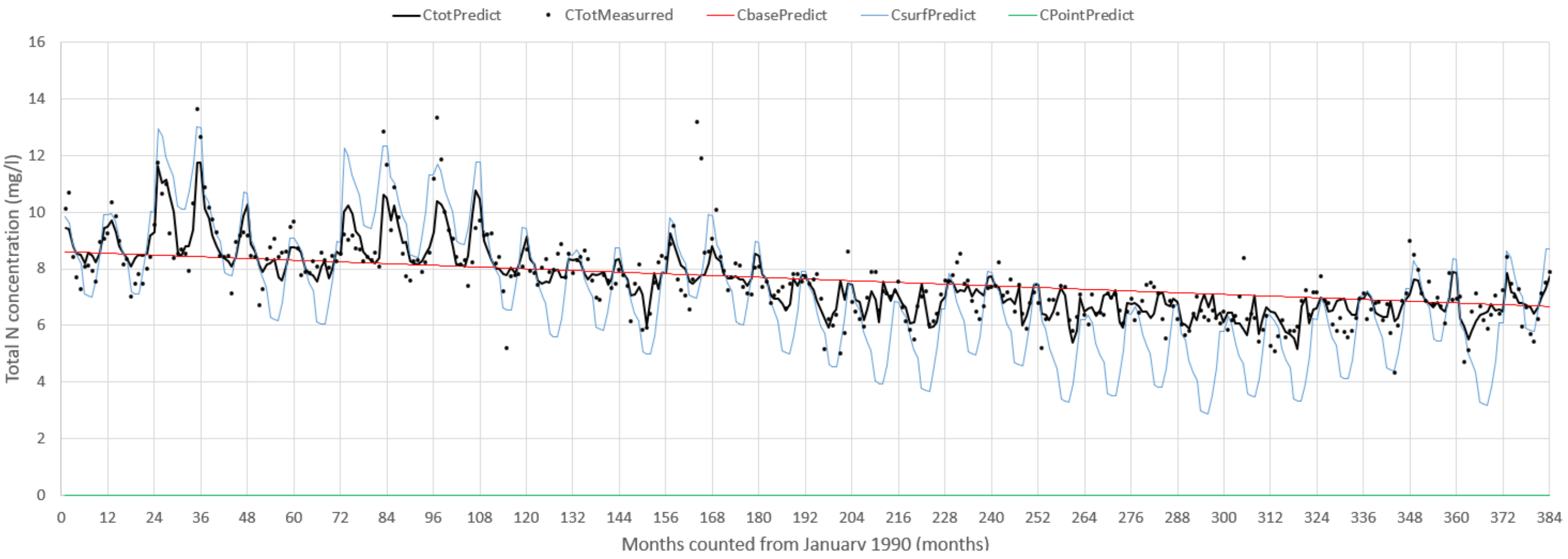

Figure 2. A time series of predicted (solid black curve) and measured inlet concentrations (circles) for N. The predicted surface water concentration is shown as a blue line, and the base-flow concentration as a red line.

The values from Table 3 are used in Eq. 25 and 26 to predicts respectively $Cin_{meabase}|_{i,j}$ and $Cin_{measurf}\big|_{i,j}$. The seasonal relation is shown in Figure 2 for 1990. $Cin_{meabase}|_{i,j}$ does not change much during one single year, so the dashed curve of $Cin_{meabase}|_{i,j}$ is nearly constant in Figure 2. The N concentration for $Cin_{measurf}\big|_{i,j}$ is likely to show variations during the year due to the effects of the growing season, having increased plant uptake in the spring and summer season leading to a drop in the $Cin_{measurf}\big|_{i,j}$ value.

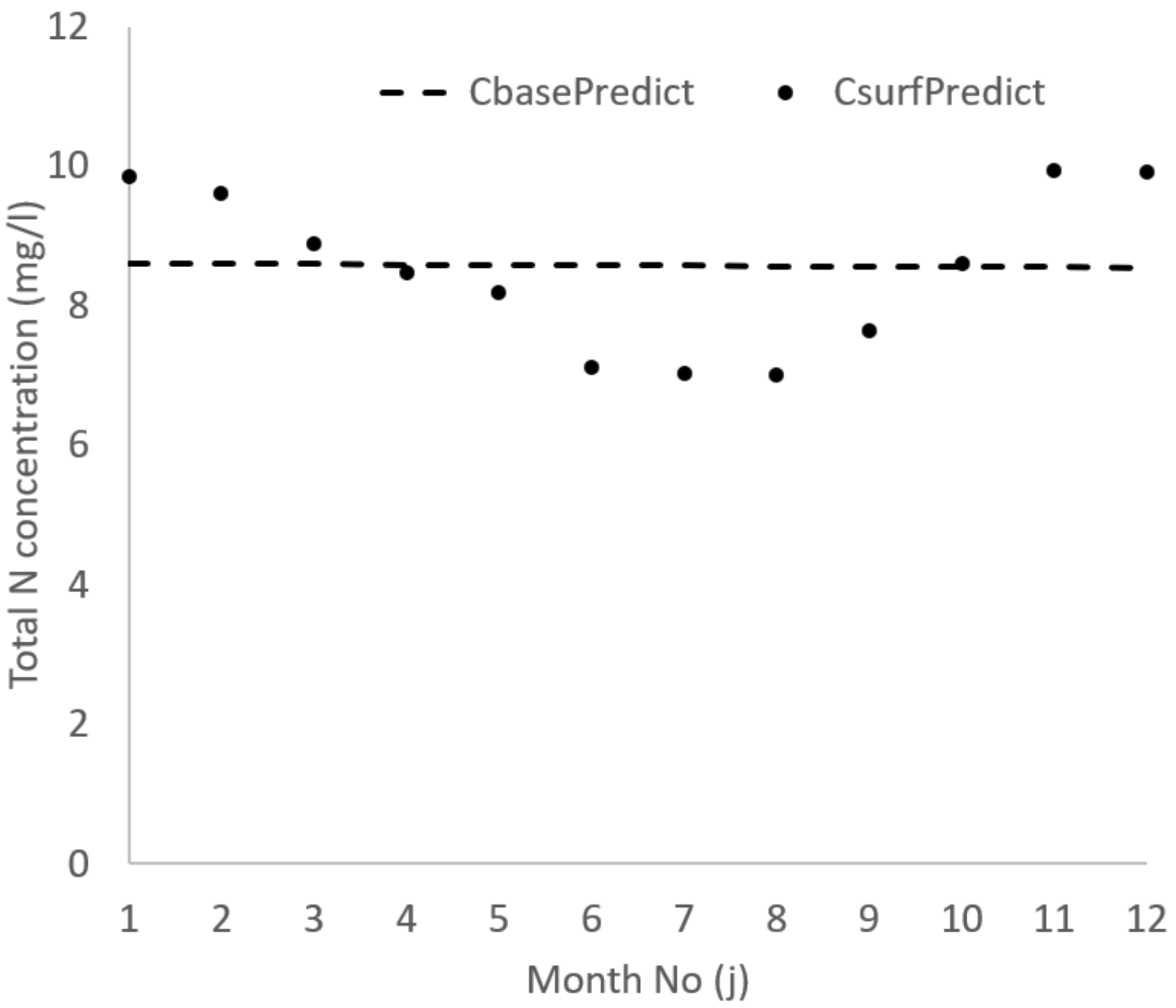

Figure 3. The variation of predicted concentration values during 1990, where CbasePredict is the predicted base water concentration, and CsurfPredict is the predicted surface-near water concentration.

The long-term variation in concentration levels is shown in Figure 3 for the winter flow concentration (January, $j$=1). The base-flow concentration shows a weak decrease during the period, while the surface-near concentration primarily shows a rapid decrease in the period 2000-2005 from a relatively high concentration level to a lower level.

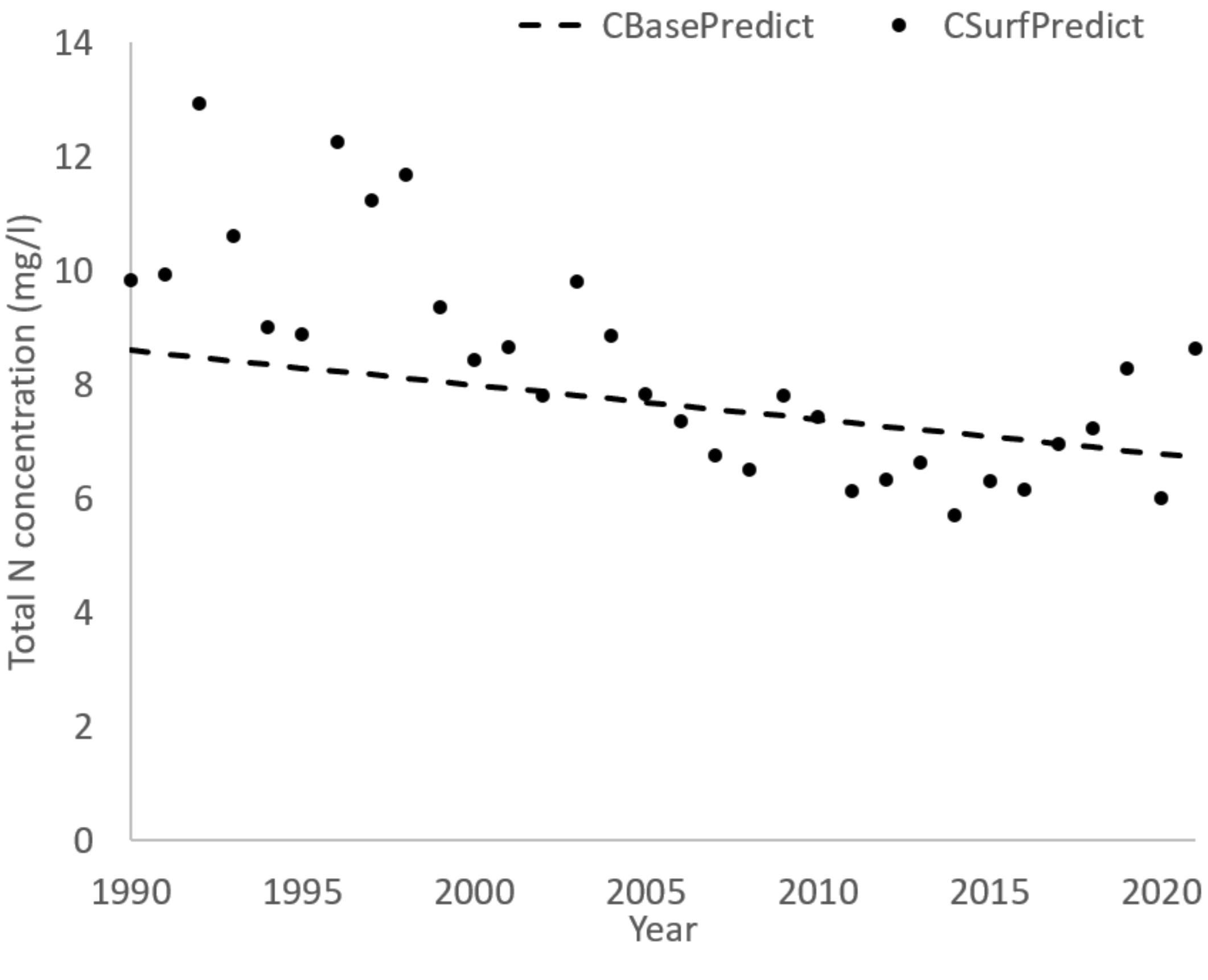

Figure 4. Long-term time trend of the N concentration (January every year), where CbasePredict is the predicted base water concentration and CsurfPredict is the predicted surface-near water concentration.

The purpose of the statistical model is primarily to predict the diffuse source load of N due to unmeasured fractions of water bodies, however, the application of the model can be evaluated by considering the capability of the model to predict the load for the measured fraction of the water bodies. Such an evaluation is shown in Figure 5.

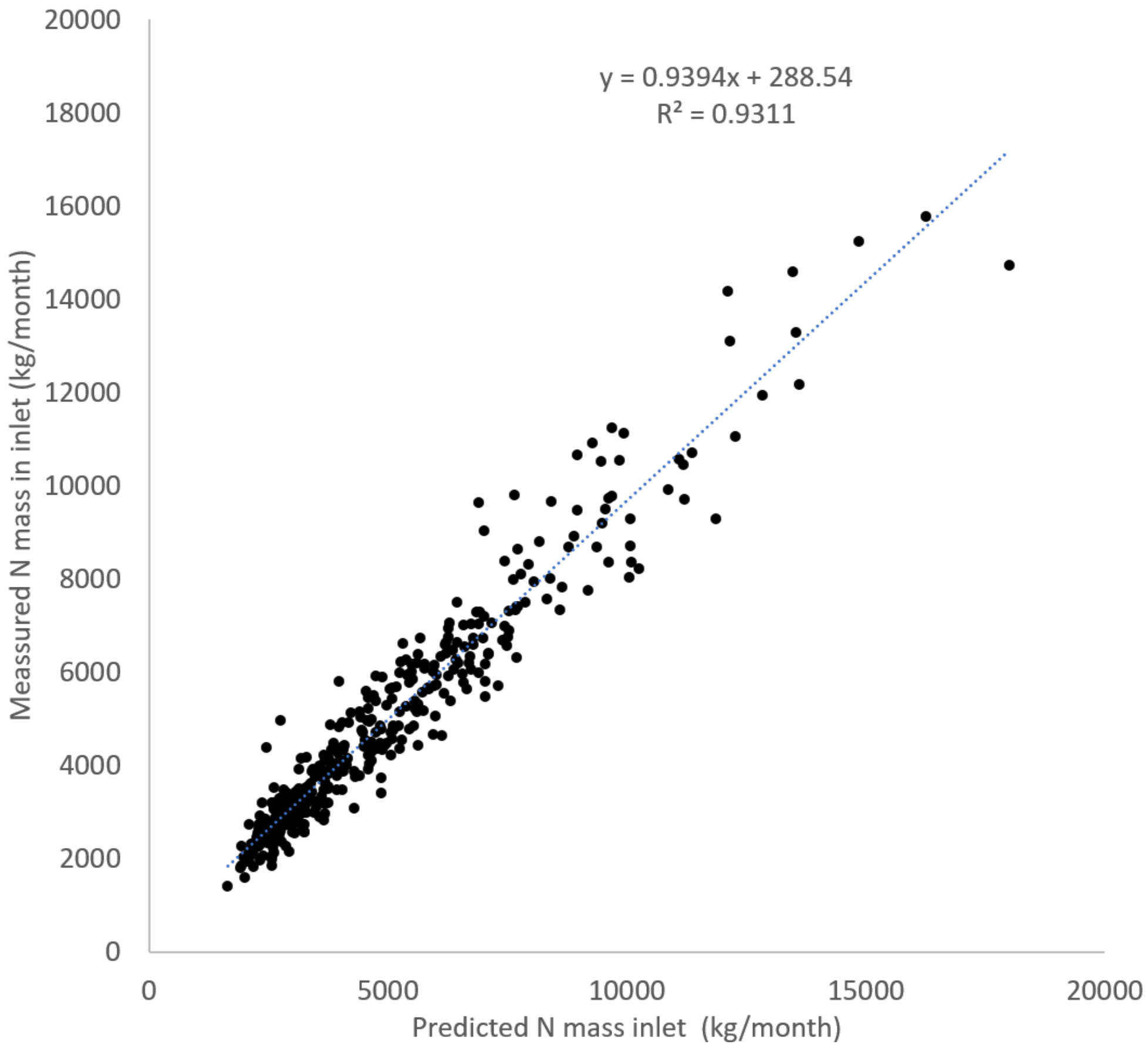

Figure 5. Comparing the predicted diffuse load of N with the measured diffuse load.

The total load in and out of the lake is shown in Figure 5. The final result for N is the predicted input and output of N for Bryrup Langsø, as shown in Figure 6, where the measured load is added to the predicted unmeasured load for each month in the time series. The inlet load is seen to be substantially larger than the outlet, disclosing a retention of N in the lake as expected. On average, the lake is removes 2500 kg N from the fresh water ecosystem every month.

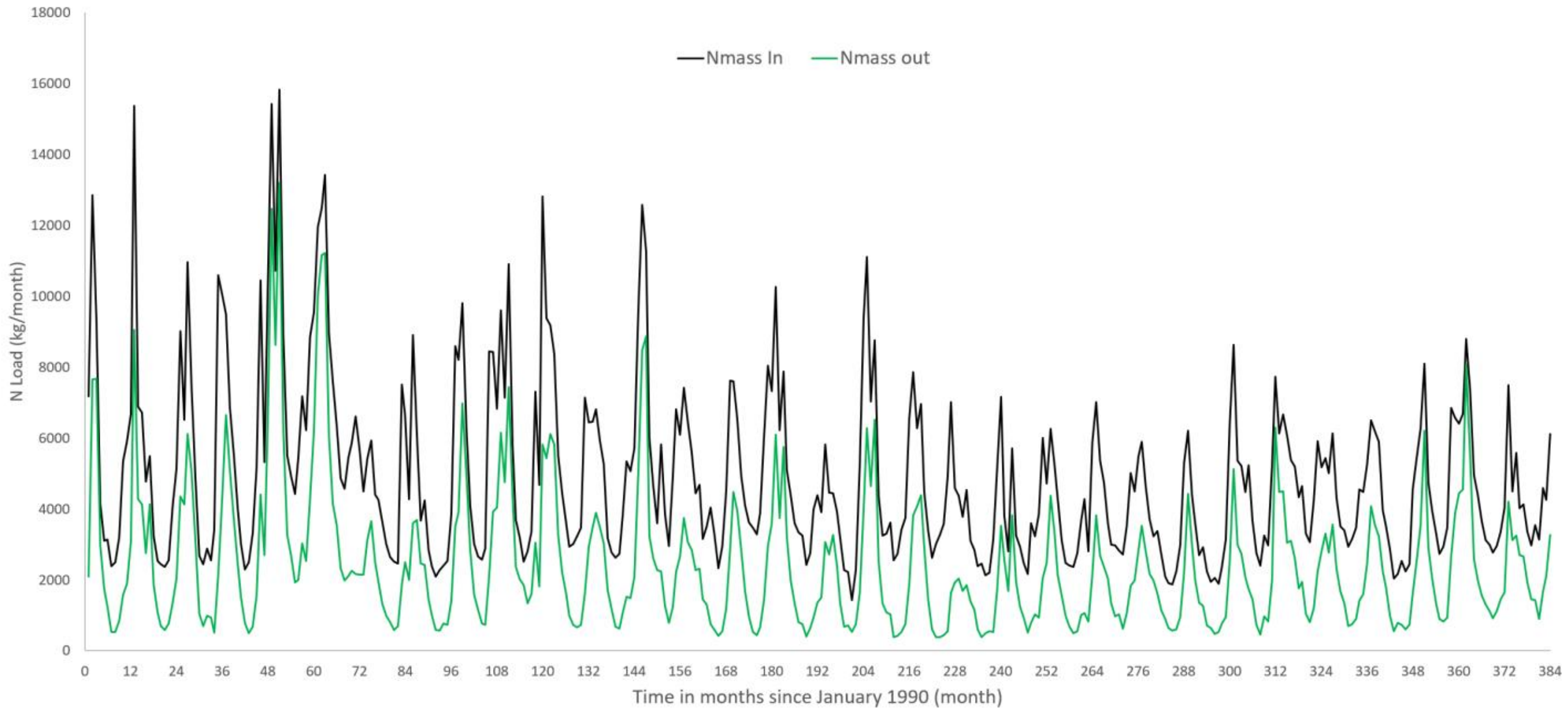

Figure 6. The total predicted load of N in and out of the Bryrup Langsø.

## P Model

The P concentration in the inlet is analyzed using the equations 24-26. The coefficient table for Eq. 24 is shown in Table 5. The point source factor in Table 5 is estimated to be 0.1056, which indicates some influence due to point sources.

| coefficient | Estimate | unit | coefficient | Estimate | unit | coefficient | Estimate | unit |
|---|---|---|---|---|---|---|---|---|
| $\gamma 1_1$ | 0 | mg/l | $\gamma 1_{17}$ | 0.004 | mg/l | $\gamma 2_1$ | 0.081 | mg/l |
| $\gamma 1_2$ | -0.039 | mg/l | $\gamma 1_{18}$ | -0.016 | mg/l | $\gamma 2_2$ | 0.078 | mg/l |
| $\gamma 1_3$ | -0.024 | mg/l | $\gamma 1_{19}$ | -0.065 | mg/l | $\gamma 2_3$ | 0.068 | mg/l |
| $\gamma 1_4$ | -0.021 | mg/l | $\gamma 1_{20}$ | -0.072 | mg/l | $\gamma 2_4$ | 0.040 | mg/l |
| $\gamma 1_5$ | -0.044 | mg/l | $\gamma 1_{21}$ | -0.039 | mg/l | $\gamma 2_5$ | 0.054 | mg/l |
| $\gamma 1_6$ | -0.093 | mg/l | $\gamma 1_{22}$ | -0.035 | mg/l | $\gamma 2_6$ | 0.085 | mg/l |
| $\gamma 1_7$ | -0.023 | mg/l | $\gamma 1_{23}$ | -0.068 | mg/l | $\gamma 2_7$ | 0.103 | mg/l |
| $\gamma 1_8$ | -0.095 | mg/l | $\gamma 1_{24}$ | -0.080 | mg/l | $\gamma 2_8$ | 0.130 | mg/l |
| $\gamma 1_9$ | -0.007 | mg/l | $\gamma 1_{25}$ | -0.069 | mg/l | $\gamma 2_9$ | 0.109 | mg/l |
| $\gamma 1_{10}$ | -0.053 | mg/l | $\gamma 1_{26}$ | -0.049 | mg/l | $\gamma 2_{10}$ | 0.099 | mg/l |
| $\gamma 1_{11}$ | -0.074 | mg/l | $\gamma 1_{27}$ | -0.064 | mg/l | $\gamma 2_{11}$ | 0.080 | mg/l |
| $\gamma 1_{12}$ | -0.003 | mg/l | $\gamma 1_{28}$ | -0.065 | mg/l | $\gamma 2_{12}$ | 0.074 | mg/l |
| $\gamma 1_{13}$ | -0.060 | mg/l | $\gamma 1_{29}$ | -0.079 | mg/l | $\theta o$ | 0.09647 | mg/l |
| $\gamma 1_{14}$ | -0.088 | mg/l | $\gamma 1_{30}$ | -0.058 | mg/l | $\theta 1$ | -0.00011 | mg/(l·t) |
| $\gamma 1_{15}$ | -0.042 | mg/l | $\gamma 1_{31}$ | -0.037 | mg/l | $\rho$ | 0.016 | - |
| $\gamma 1_{16}$ | -0.049 | mg/l | $\gamma 1_{32}$ | -0.068 | | | | |

Table 5. Estimated parameters for Eq. (23), $R^2$=0.49.

The time series of predicted concentration is show in Figure 7 together with the measured values subtracted point source estimates. Some extreme values, especially some higher levels, seem poorly described by the model.

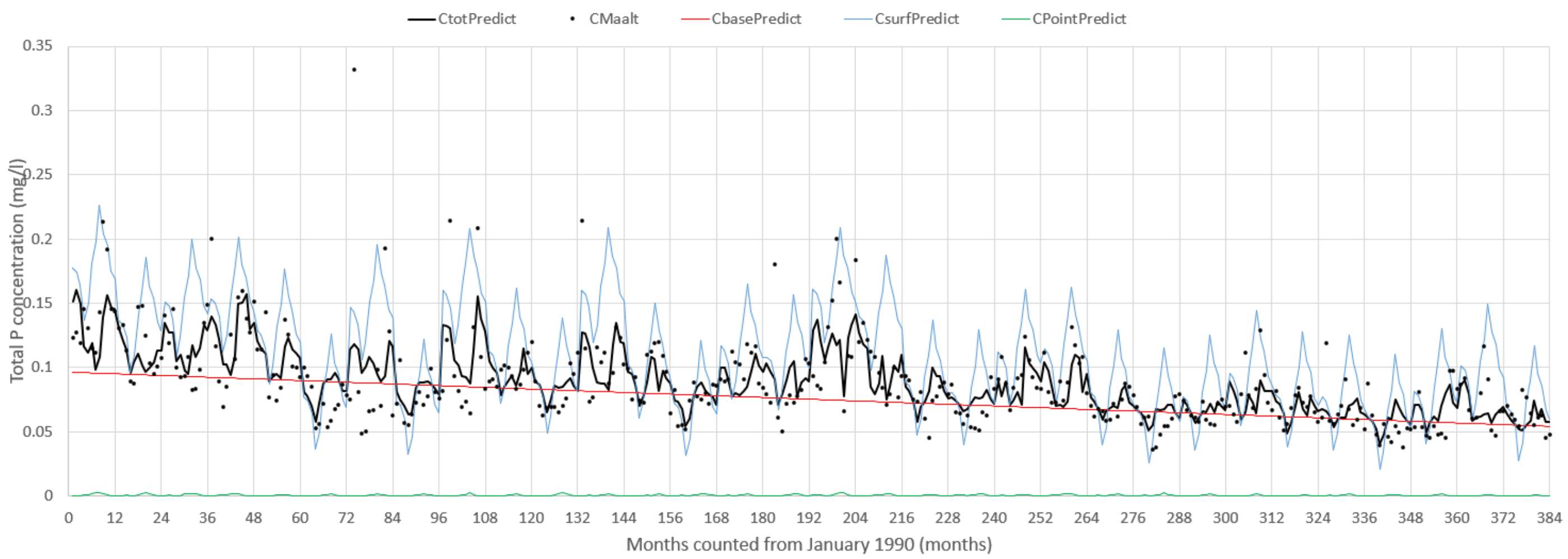

Figure 7. A time series of predicted (solid black curve) and measured inlet concentrations (circles) of P. The predicted surface water concentration is shown as a blue line and the base-flow concentration as a red line.

The values from Table 5 are used in Eqs. 24 and 25 to predicts respectively $Cin_{meabase}|_{i,j}$ and $Cin_{measurf}\big|_{i,j}$. The seasonal relation is shown in Figure 2 for 1990. $Cin_{meabase}|_{i,j}$ does not change much during one single year, so the dashed curve of $Cin_{meabase}|_{i,j}$ is nearly constant in Figure 2. The P concentration for $Cin_{measurf}\big|_{i,j}$ is not likely to show variations during the year, as the N concentration in Figure 3 indicates another fate and transport mechanism for P compared to N, where P is more likely to be governed by adsorption kinetic and not so much by plant uptake as N.

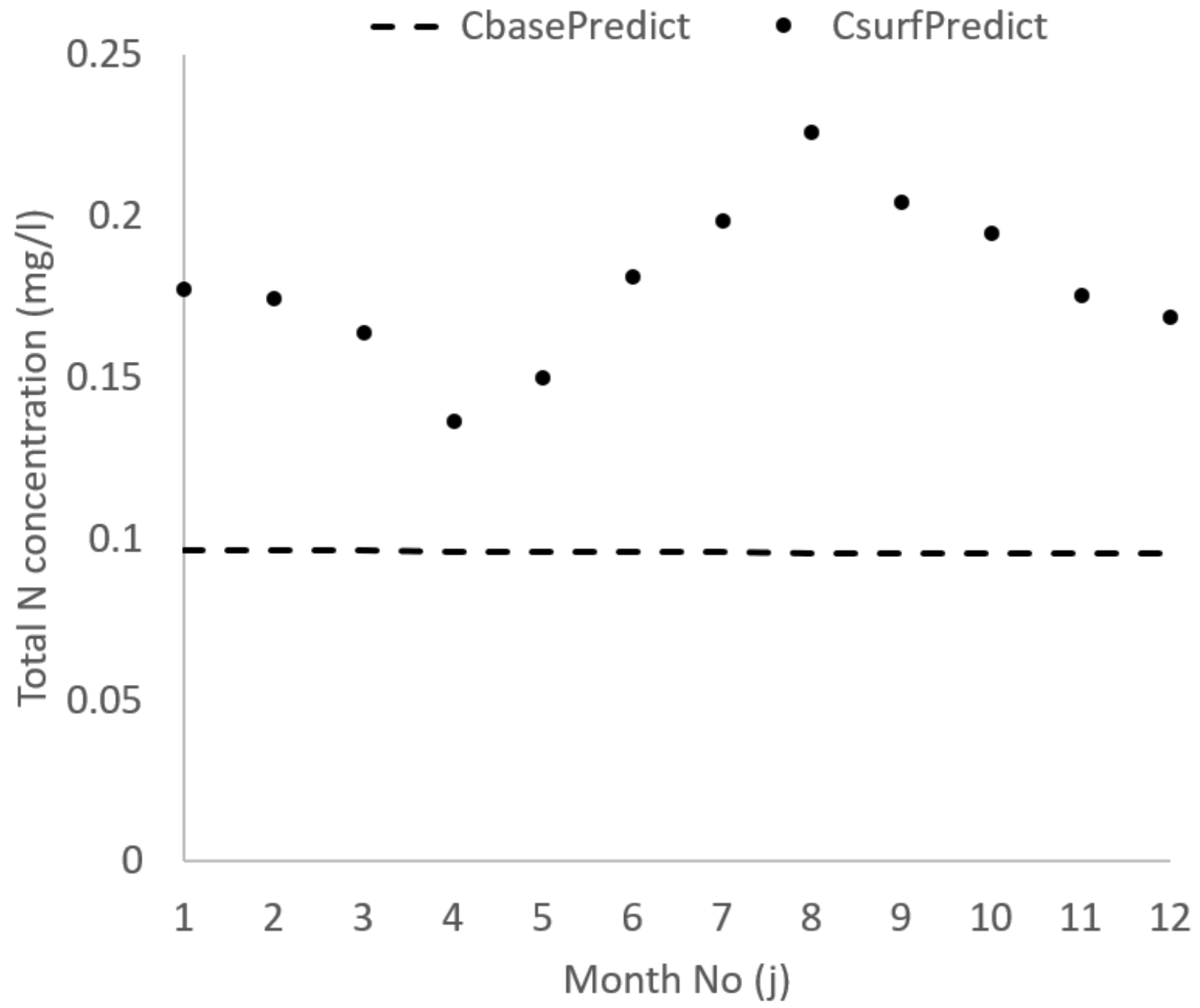

Figure 8. The variation of predicted concentration values during year 1990, where CbasePredict is the predicted base water concentration and CsurfPredict is the predicted surface-near water concentration.

The long-term variation in concentration levels is shown for P in Figure 9 for the winter flow concentration (January, *j*=1). The surface concentration of P is more likely to show a continuous drop during the entire period compared to the similar graph for N (Figure 4).

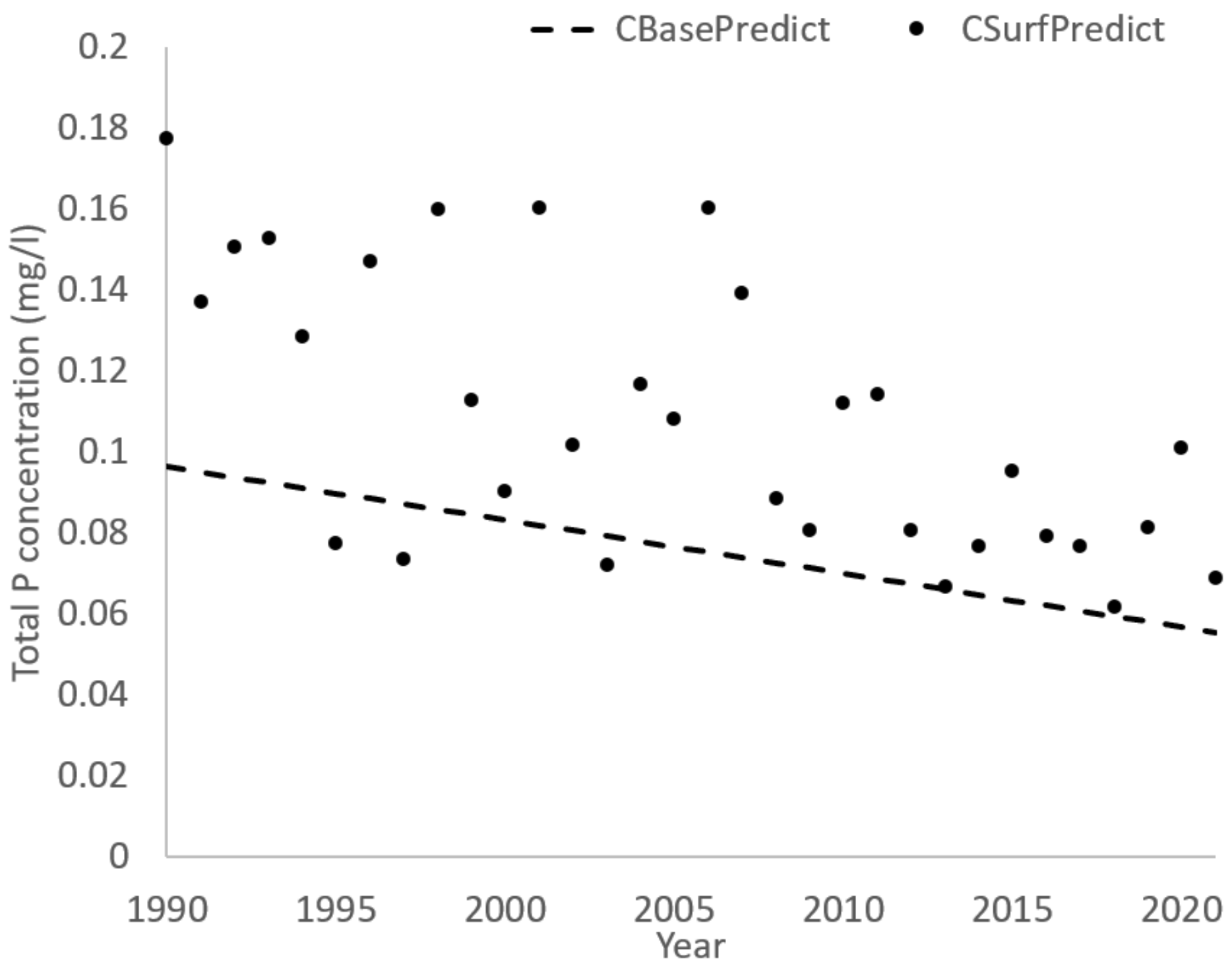

Figure 9. Long-term time trend of the N concentration (January), where CbasePredict is the predicted base water concentration and CsurfPredict is the predicted surface-near water concentration.

The load of P from the measured part of the lake catchment is compared with the measured load in Figure 10.

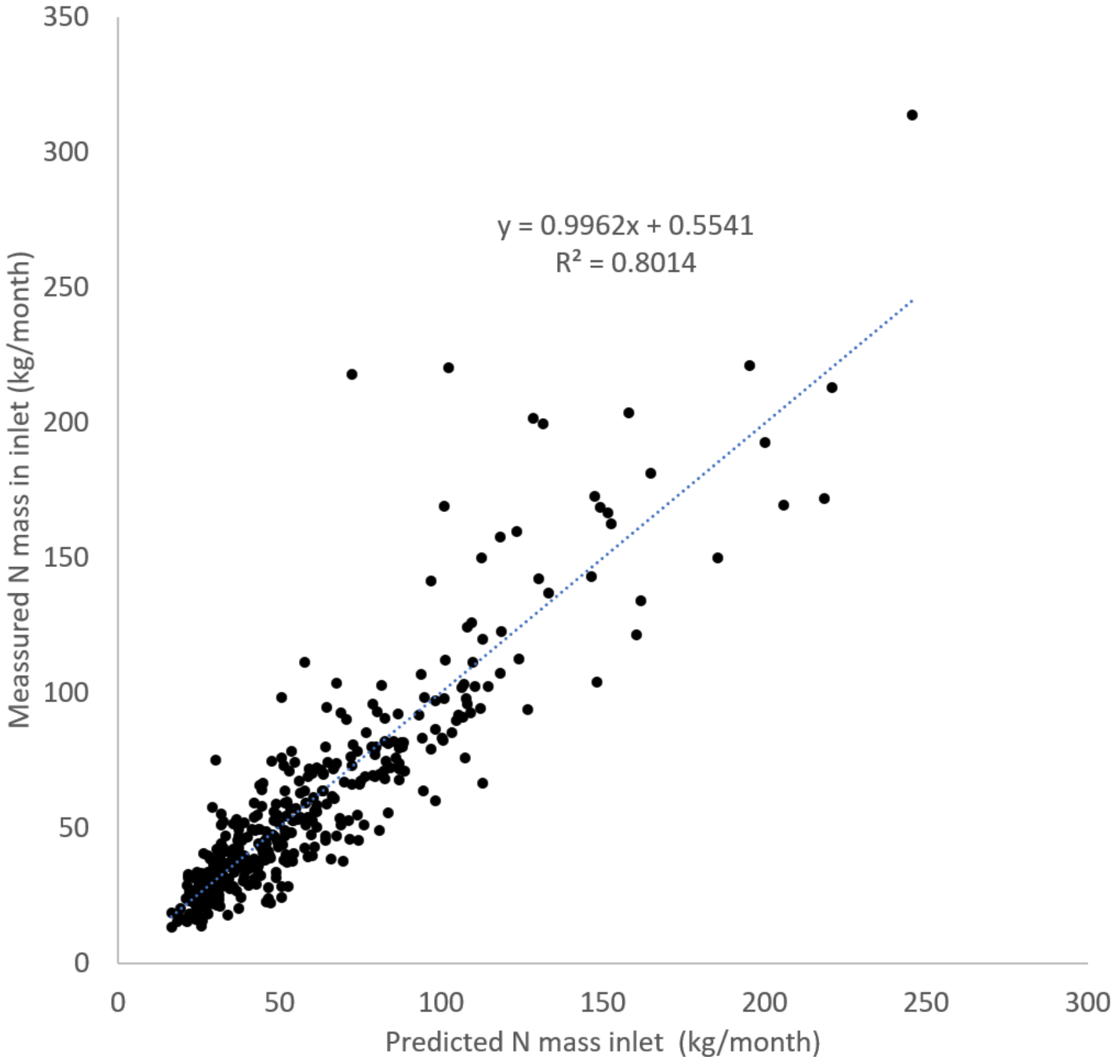

Figure 10. Comparison of the predicted load of P with the measured diffuse load.

The total load in an out of the lake is shown in Figure 11. The retention of P in the lake is not so pronounced as for N. The average retention is 14 kg/month.

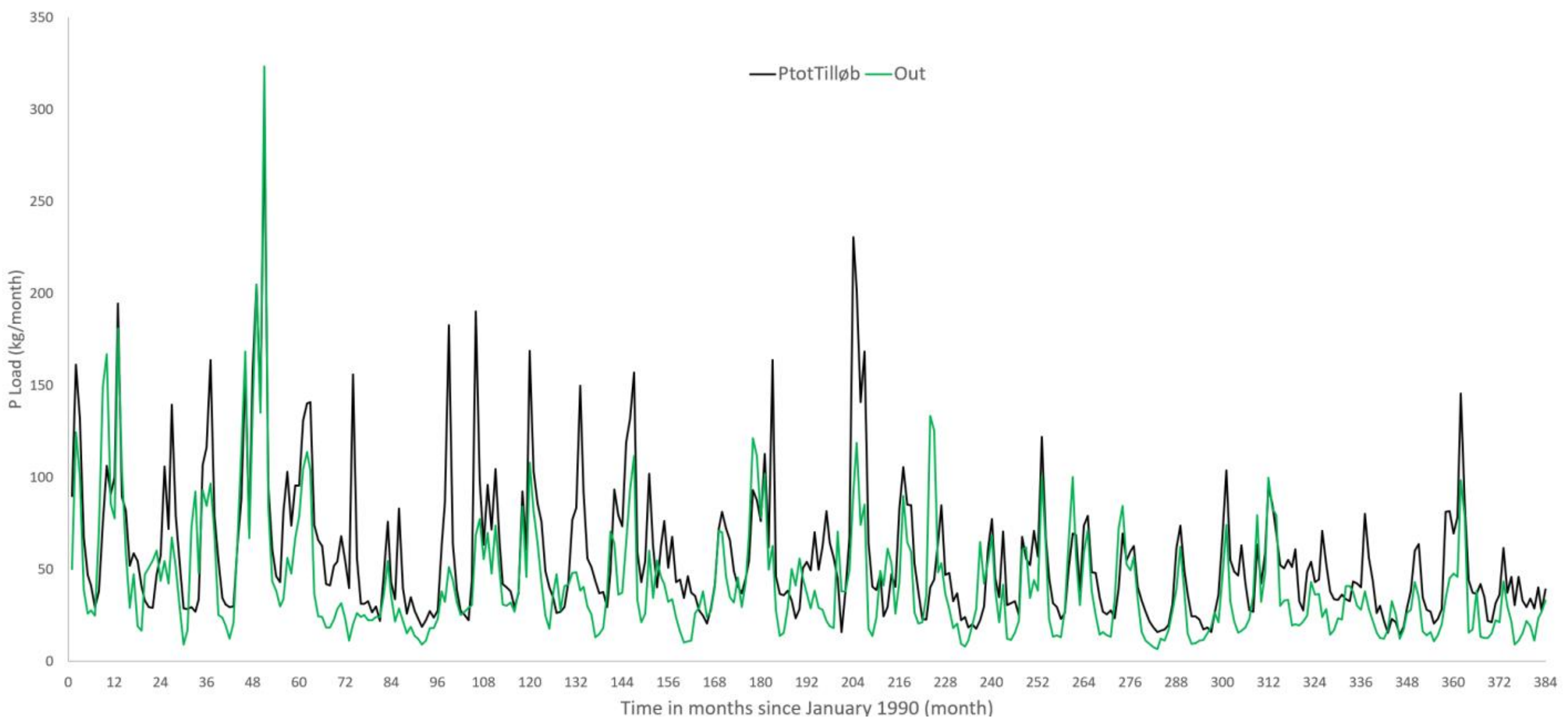

Figure 11. The total load of P in and out of the Bryrup Langsø.

## Discussion

The purpose of the development of the presented statistical model concept is to predict the load of N and P in and out of a lake, where only a fraction of the inflow and outflow is measured, and the unmeasured fraction thus needs to be predicted. In most cases, the lake will only have one single outlet, so the outflow tends to be better described than the inflow. However, lakes may have seepage from the bottom that will contribute to the release of N and P in the same way as the outlet at the surface. The inlet to the lake may consist of several rivers and streams, out of which only a fraction of the water body is measured. There may be a main tributary entering the lake, while smaller streams are unmeasured. Obviously, there are uncertainties related to the measured load of both water and nutrient concentration levels that will alter the load prediction also for the measured fraction of the water bodies. However, in this method the focus has solely been allocated to predicting the unmeasured fraction of the load. Thus, proper quality assurance of the monitoring programs that are used as data input is critical for the load calculation. The well-known general modelling condition of “garbage in is garbage out” will also apply for the load predictions in case the data input to the load predictions is not of sufficient quality and, thus, introduces significant uncertainty.

The model assumes that the minimum monthly measured inlet flow of water during a year is equivalent to the base-flow. Thus, the model assumes a seasonality of the discharge, where there is a dry season and/or a season where the evaporation from the soil and plant surfaces evaporates all the precipitation. Better models could be applied for estimating the base-flow for the reference month. Extreme rain events during summer may lead to overestimation of the base-flow during the reference month.

Conceptually, the model relies on a linear mass balance model relating measured outflow to measured inflow. The model coefficients are used to predict the unmeasured part separated into a baseflow and a surface-near flow. This separation is beneficial from the point of view that the concentration levels of N and P not necessarily are similar. The intercept of the linear model can be interpreted as the volume of water that runs out of the lake when no measured water enters the lake. Thus, the estimated intercept predicts the water exchange through the lake bottom or from unmeasured wells entering the lake, and a negative value estimate indicates loss of water due to seepage through the lake bottom. The slope between inflow and outflow is assumed to reflect the fraction of unmeasured water in the inflowing rivers and streams. This is based on the rationale that if e.g. 2 $m^3$ water is measured at the outlet of the lake for every measured 1 $m^3$ at the inlet (slope equal 2), then there is 1 $m^3$ unmeasured water in the inflow for every 1 $m^3$ of measured water volume. For the baseflow, the assumption is that the change in intercept between months is equivalent to a similar change in measured baseflow. This is combined with the assumption that the smallest measured monthly inlet flow of water is entirely baseflow.

The accumulation of water is included as a memory effect in a linear model that includes last month's water input to model the actual month's output. Thus, the water accumulation impact on water outflow is assumed to be linear between the incoming volume and the resulting outgoing volume the month after. This is valid for smaller charges in water volume, as the larger changes in the relation between incoming water volume and outgoing water volume the month after tends to be nonlinear.

Any replacement of the measuring stations is assumed to take place at the start of a year. However, the statistical model for the water balance could be modified to include changes in measuring starting at any month and year.

The seasonality in base-flow is assumed to be independent of the measurement station combination, and this assumption could be challenged in case a large amount of unmeasured water exists. However, in case the amount of unmeasured water is large, it will always be a challenge to make valid load prediction.

The N and P concentrations are predicted using a statistical model that decomposes the sources of nutrient into three components: (1) point sources; (2) surface-near sources; (3) base-flow source. The point source's strength must be identified both in the catchment of the measured inflow and in the catchment of the lake, and the model will estimate the effective load of the point source strength. However, some fraction of the point source contribution may be modelled as part of the baseflow, if the point sources are located at longer distance upstream from the lake, and especially if there are other lakes upstream that influence the concentration level of the inflowing water. The model works best for lakes that have inflow from streams and rivers not having other lakes upstream. In case of lake systems where water from one lake flows into other lakes downstream, a more comprehensive model for the load in and out each lake may be better. Such a model could apply the model in this paper for the upper catchment lakes using the principle in the paper.

The separation of the water body into respectively baseflow and surface-near flow is a rough categorization used to include the large variation in water transport routes. However, the specific case where the surface-near concentration level makes a decrease during the growing season indicates some realism behind the model. The surface-near sources of N and P are considered to come from the upper soil layer or from run off and drain systems in the catchment and will contribute to the surface-near flow as well.

Lake load of N and P was modelled using similar monitoring data from NOVANA by Windolf et al., (1996). Their approach was different from the approach in the paper in relation to:

- The lake water volume needed to be determined for each month to solve the mass balance.
- The mass balance of water was more aggregated without distinction between surface-near and base-flow water, which can have the consequence that unmeasured surface-near water may be hidden in seepage or ground water inflow through the lake bottom in the mass balance equation.
- The inlet concentration was not decomposed in surface-near water concentration and base-flow concentration.
- The unmeasured base-flow concentration needed to be estimated using additional information about ground water concentration levels.

In some circumstances, it may be critical that the method in this paper uses the predicted base-flow concentration to predict the unmeasured base-flow concentration in cases where point sources may be hidden as base-flow concentration due to upstream lakes. In such cases, the method can be adjusted by replacing the predicted base-flow concentration with independent measurements of ground water concentrations, as done by Jørgen et al., 1996.

# Appendix A

## Table A1

The table is also available as supplementary material.

| Year | i | j | $Qout_{mea}\|_{i,j}$ | Station location | $Qin_{mea}\|_{i,j}$ | $jref_i$ | $Qin_{vertical}\|_{i,j}$ | N | | | | P | | | |
|---|---|---|---|---|---|---|---|---|---|---|---|---|---|---|---|
| | | | | | | | | $CNin_{totmea}\|_{i,j}$ | $CNout_{totmea}\|_{i,j}$ | $P_{mea}\|_{i,j}$ | $P_{unmea}\|_{i,j}$ | $CPin_{totmea}\|_{i,j}$ | $CPout_{totmea}\|_{i,j}$ | $P_{mea}\|_{i,j}$ | $P_{unmea}\|_{i,j}$ |
| | | | $m^3$/month | | $m^3$/month | | $m^3$/month | mg/l | mg/l | Kg/month | Kg/month | mg/l | mg/l | Kg/month | Kg/month |
| 1990 | 1 | 1 | 541037 | 1 | 545773 | 7 | 46169 | 10.109 | 3.871 | 121 | 79.4 | 0.1231 | 0.0932 | 19.3 | 12.9 |
| 1990 | 1 | 2 | 1328141 | 1 | 958569 | 7 | 47883 | 10.672 | 5.767 | 145 | 80.1 | 0.1276 | 0.0938 | 25.7 | 13.0 |
| 1990 | 1 | 3 | 1127606 | 1 | 811620 | 7 | 10843 | 8.430 | 6.800 | 165 | 77.3 | 0.1189 | 0.0897 | 30.2 | 12.6 |
| 1990 | 1 | 4 | 492480 | 1 | 344308 | 7 | -6186 | 7.716 | 6.024 | 119 | 71.6 | 0.1457 | 0.0789 | 18.6 | 11.7 |
| 1990 | 1 | 5 | 374976 | 1 | 244786 | 7 | -26419 | 7.290 | 4.661 | 107 | 68.6 | 0.1309 | 0.0698 | 16.4 | 11.6 |
| 1990 | 1 | 6 | 375840 | 1 | 239664 | 7 | 3130 | 8.070 | 3.323 | 135 | 66.9 | 0.1145 | 0.0736 | 25.7 | 11.2 |
| 1990 | 1 | 7 | 291946 | 1 | 179378 | 7 | -23811 | 8.098 | 1.826 | 139 | 67.3 | 0.1115 | 0.0856 | 26.6 | 11.1 |
| 1990 | 1 | 8 | 294624 | 1 | 193992 | 7 | -1379 | 7.912 | 1.796 | 174 | 68.0 | 0.1431 | 0.2409 | 36.2 | 11.2 |
| 1990 | 1 | 9 | 520992 | 1 | 290387 | 7 | 47547 | 7.539 | 1.609 | 155 | 68.0 | 0.2133 | 0.2865 | 29.3 | 10.9 |
| 1990 | 1 | 10 | 739238 | 1 | 459224 | 7 | 33350 | 8.948 | 2.124 | 135 | 69.8 | 0.1915 | 0.2258 | 22.8 | 11.2 |
| 1990 | 1 | 11 | 593568 | 1 | 491495 | 7 | 18706 | 9.063 | 3.177 | 107 | 72.4 | 0.1456 | 0.1446 | 15.2 | 11.5 |
| 1990 | 1 | 12 | 666922 | 1 | 550819 | 7 | 25898 | 9.265 | 4.615 | 114 | 73.9 | 0.1438 | 0.1168 | 17.6 | 11.6 |
| 1991 | 2 | 1 | 1473120 | 1 | 1192849 | 8 | 35362 | 10.347 | 6.139 | 85 | 74.6 | 0.1308 | 0.1228 | 12.2 | 20.9 |
| 1991 | 2 | 2 | 613993 | 1 | 508217 | 8 | 14570 | 9.862 | 6.955 | 109 | 75.4 | 0.1331 | 0.1717 | 18.7 | 21.0 |
| 1991 | 2 | 3 | 703616 | 1 | 507748 | 8 | 9204 | 8.999 | 5.871 | 127 | 72.6 | 0.1131 | 0.0848 | 23.2 | 20.6 |
| 1991 | 2 | 4 | 536803 | 1 | 380042 | 8 | 14644 | 8.145 | 5.120 | 83 | 66.8 | 0.0895 | 0.0546 | 11.8 | 19.7 |
| 1991 | 2 | 5 | 767629 | 1 | 445694 | 8 | -18333 | 8.330 | 5.379 | 73 | 63.9 | 0.0874 | 0.0617 | 9.5 | 19.6 |
| 1991 | 2 | 6 | 452304 | 1 | 267244 | 8 | 3838 | 7.007 | 4.100 | 102 | 62.1 | 0.1470 | 0.0426 | 18.5 | 19.2 |
| 1991 | 2 | 7 | 344174 | 1 | 198789 | 8 | -26009 | 7.478 | 3.116 | 106 | 62.5 | 0.1480 | 0.0492 | 19.5 | 19.1 |
| 1991 | 2 | 8 | 287928 | 1 | 181390 | 8 | -16060 | 7.815 | 2.423 | 142 | 63.2 | 0.1248 | 0.1654 | 29.4 | 19.2 |
| 1991 | 2 | 9 | 287194 | 1 | 192924 | 8 | 1155 | 7.462 | 2.047 | 121 | 63.2 | 0.1034 | 0.1778 | 22.7 | 18.9 |
| 1991 | 2 | 10 | 381672 | 1 | 212856 | 8 | 16433 | 7.990 | 2.035 | 100 | 65.0 | 0.0947 | 0.1442 | 16.1 | 19.1 |
| 1991 | 2 | 11 | 565574 | 1 | 335956 | 8 | 37784 | 8.432 | 2.376 | 71 | 67.6 | 0.1010 | 0.1066 | 8.3 | 19.5 |
| 1991 | 2 | 12 | 595408 | 1 | 397871 | 8 | 26345 | 9.539 | 3.431 | 78 | 69.2 | 0.1070 | 0.0736 | 10.4 | 19.6 |
| 1992 | 3 | 1 | 894586 | 1 | 596288 | 7 | 24556 | 11.765 | 4.861 | 95 | 48.4 | 0.1408 | 0.0610 | 12.4 | 12.0 |
| 1992 | 3 | 2 | 688288 | 1 | 443764 | 7 | 16917 | 10.660 | 5.993 | 119 | 49.7 | 0.1187 | 0.0614 | 18.9 | 12.1 |
| 1992 | 3 | 3 | 976009 | 1 | 748829 | 7 | 29177 | 10.975 | 6.256 | 138 | 45.2 | 0.1459 | 0.0691 | 23.4 | 11.5 |
| 1992 | 3 | 4 | 818294 | 1 | 581042 | 7 | 16843 | 9.243 | 6.187 | 93 | 36.1 | 0.1001 | 0.0631 | 12.0 | 10.1 |
| 1992 | 3 | 5 | 570767 | 1 | 386877 | 7 | -25674 | 8.377 | 5.502 | 83 | 31.4 | 0.0923 | 0.0519 | 9.7 | 9.9 |
| 1992 | 3 | 6 | 242352 | 1 | 183332 | 7 | -49709 | 8.515 | 4.328 | 111 | 28.6 | 0.0930 | 0.0375 | 18.8 | 9.2 |
| 1992 | 3 | 7 | 227128 | 1 | 175485 | 7 | -19936 | 8.675 | 3.030 | 115 | 29.3 | 0.1082 | 0.0760 | 19.7 | 9.1 |
| 1992 | 3 | 8 | 366673 | 1 | 221493 | 7 | 31375 | 8.519 | 2.682 | 151 | 30.4 | 0.0829 | 0.1999 | 29.6 | 9.2 |
| 1992 | 3 | 9 | 362102 | 1 | 209606 | 7 | -3056 | 7.941 | 2.576 | 131 | 30.4 | 0.0838 | 0.2556 | 22.9 | 8.8 |
| 1992 | 3 | 10 | 315783 | 1 | 245744 | 7 | 20047 | 10.304 | 1.632 | 110 | 33.2 | 0.0981 | 0.1524 | 16.3 | 9.2 |
| 1992 | 3 | 11 | 825552 | 1 | 630509 | 7 | 50417 | 13.628 | 2.569 | 81 | 37.4 | 0.1348 | 0.1129 | 8.5 | 9.8 |
| 1992 | 3 | 12 | 833518 | 1 | 636757 | 7 | 19563 | 12.648 | 5.086 | 88 | 39.8 | 0.1490 | 0.1015 | 10.7 | 9.9 |
| 1993 | 4 | 1 | 1005757 | 1 | 691131 | 6 | 48106 | 10.866 | 6.614 | 105 | 48.7 | 0.2005 | 0.0963 | 12.6 | 12.2 |
| 1993 | 4 | 2 | 716910 | 1 | 502181 | 6 | 11589 | 10.171 | 7.089 | 130 | 49.9 | 0.1162 | 0.0964 | 19.1 | 12.4 |
| 1993 | 4 | 3 | 571184 | 1 | 390190 | 6 | -633 | 9.758 | 6.525 | 149 | 45.5 | 0.0892 | 0.0445 | 23.6 | 11.7 |
| 1993 | 4 | 4 | 417624 | 1 | 287758 | 6 | -14085 | 9.270 | 5.892 | 104 | 36.3 | 0.0697 | 0.0570 | 12.2 | 10.3 |
| 1993 | 4 | 5 | 324421 | 1 | 211479 | 6 | -28282 | 8.467 | 4.631 | 92 | 31.5 | 0.0849 | 0.0610 | 9.9 | 10.1 |
| 1993 | 4 | 6 | 252008 | 1 | 151782 | 6 | -30369 | 8.416 | 3.111 | 121 | 28.7 | 0.1260 | 0.0501 | 19.0 | 9.4 |
| 1993 | 4 | 7 | 300487 | 1 | 188426 | 6 | 9986 | 8.469 | 1.641 | 125 | 29.4 | 0.1063 | 0.0695 | 19.9 | 9.3 |
| 1993 | 4 | 8 | 469956 | 1 | 311278 | 6 | 21128 | 7.129 | 1.423 | 160 | 30.5 | 0.1550 | 0.1385 | 29.8 | 9.4 |
| 1993 | 4 | 9 | 671221 | 1 | 441968 | 6 | 35437 | 8.933 | 2.336 | 141 | 30.5 | 0.1594 | 0.1710 | 23.1 | 8.9 |
| 1993 | 4 | 10 | 1082584 | 1 | 911637 | 6 | 32866 | 9.162 | 4.072 | 120 | 33.4 | 0.1384 | 0.1555 | 16.5 | 9.3 |
| 1993 | 4 | 11 | 545317 | 1 | 427990 | 6 | 19004 | 9.290 | 4.936 | 92 | 37.6 | 0.1276 | 0.1229 | 8.7 | 9.9 |
| 1993 | 4 | 12 | 1140036 | 1 | 853109 | 6 | 63794 | 9.168 | 5.565 | 98 | 40.0 | 0.1514 | 0.1187 | 10.9 | 10.0 |
| 1994 | 5 | 1 | 1864682 | 1 | 1402574 | 7 | 51870 | 8.459 | 6.683 | 96 | 48.7 | 0.1142 | 0.1099 | 12.2 | 9.2 |
| 1994 | 5 | 2 | 1256181 | 1 | 917748 | 7 | 34170 | 8.422 | 6.871 | 124 | 49.9 | 0.1137 | 0.1078 | 19.9 | 9.3 |
| 1994 | 5 | 3 | 2190389 | 1 | 1666050 | 7 | 39797 | 6.729 | 6.026 | 146 | 45.5 | 0.1433 | 0.1477 | 25.2 | 8.8 |
| 1994 | 5 | 4 | 1191405 | 1 | 827128 | 7 | -4732 | 7.293 | 5.210 | 94 | 36.3 | 0.0769 | 0.0709 | 11.9 | 7.8 |
| 1994 | 5 | 5 | 676302 | 1 | 394295 | 7 | -22879 | 8.359 | 4.839 | 82 | 31.5 | 0.0939 | 0.0642 | 9.2 | 7.7 |
| 1994 | 5 | 6 | 589163 | 1 | 338979 | 7 | -4024 | 8.762 | 4.558 | 116 | 28.7 | 0.0744 | 0.0654 | 19.7 | 7.2 |
| 1994 | 5 | 7 | 450664 | 1 | 312895 | 7 | -42889 | 9.062 | 4.281 | 121 | 29.4 | 0.0846 | 0.0666 | 20.8 | 7.1 |
| 1994 | 5 | 8 | 501657 | 1 | 439310 | 7 | 17625 | 8.402 | 3.972 | 163 | 30.5 | 0.1373 | 0.0678 | 32.6 | 7.2 |
| 1994 | 5 | 9 | 819980 | 1 | 624018 | 7 | 41846 | 8.556 | 3.698 | 139 | 30.5 | 0.1259 | 0.0690 | 24.8 | 6.9 |
| 1994 | 5 | 10 | 716942 | 1 | 539136 | 7 | 19637 | 8.593 | 3.522 | 113 | 33.4 | 0.1008 | 0.0666 | 17.1 | 7.2 |
| 1994 | 5 | 11 | 1019887 | 1 | 709344 | 7 | 35251 | 9.467 | 4.195 | 80 | 37.6 | 0.0998 | 0.0657 | 7.8 | 7.6 |
| 1994 | 5 | 12 | 1104460 | 1 | 758765 | 7 | 42591 | 9.661 | 5.622 | 88 | 40.0 | 0.0926 | 0.0715 | 10.2 | 7.7 |
| 1995 | 6 | 1 | 1499720 | 2 | 916218 | 8 | 37635 | 8.738 | 6.680 | 96 | 34.9 | 0.0999 | 0.0698 | 11.4 | 6.6 |
| 1995 | 6 | 2 | 1710308 | 2 | 989715 | 8 | 41585 | 7.777 | 6.524 | 117 | 35.7 | 0.0932 | 0.0666 | 17.0 | 6.7 |
| 1995 | 6 | 3 | 1854683 | 2 | 972919 | 8 | 30220 | 7.871 | 6.051 | 133 | 32.9 | 0.0850 | 0.0561 | 20.8 | 6.4 |
| 1995 | 6 | 4 | 1105723 | 2 | 587235 | 8 | -2199 | 7.851 | 5.499 | 95 | 27.2 | 0.0532 | 0.0333 | 11.1 | 5.8 |
| 1995 | 6 | 5 | 813004 | 2 | 425656 | 8 | -11850 | 8.272 | 5.097 | 85 | 24.2 | 0.0561 | 0.0302 | 9.1 | 5.7 |
| 1995 | 6 | 6 | 743810 | 2 | 345479 | 8 | -7117 | 8.089 | 4.767 | 108 | 22.5 | 0.0719 | 0.0327 | 16.9 | 5.4 |
| 1995 | 6 | 7 | 525709 | 2 | 252979 | 8 | -33238 | 8.576 | 4.424 | 112 | 22.9 | 0.0539 | 0.0352 | 17.7 | 5.4 |
| 1995 | 6 | 8 | 487273 | 2 | 228968 | 8 | -28432 | 8.302 | 4.068 | 141 | 23.6 | 0.0585 | 0.0379 | 26.0 | 5.4 |
| 1995 | 6 | 9 | 565553 | 2 | 367746 | 8 | 26755 | 8.043 | 3.730 | 126 | 23.6 | 0.0677 | 0.0404 | 20.3 | 5.2 |
| 1995 | 6 | 10 | 665665 | 2 | 426717 | 8 | 4546 | 8.469 | 3.392 | 108 | 25.4 | 0.0707 | 0.0429 | 14.7 | 5.4 |
| 1995 | 6 | 11 | 700704 | 2 | 467102 | 8 | 26531 | 8.285 | 3.085 | 85 | 28.0 | 0.0866 | 0.0451 | 8.1 | 5.7 |
| 1995 | 6 | 12 | 612749 | 2 | 368611 | 8 | 12856 | 8.540 | 3.510 | 90 | 29.5 | 0.0785 | 0.0382 | 10.0 | 5.7 |
| 1996 | 7 | 1 | 526546 | 2 | 344503 | 9 | 2832 | 9.214 | 4.068 | 81 | 35.0 | 0.0747 | 0.0214 | 9.0 | 6.7 |
| 1996 | 7 | 2 | 655597 | 2 | 414435 | 9 | 17849 | 9.028 | 4.761 | 96 | 35.8 | 0.3322 | 0.0309 | 13.0 | 6.7 |
| 1996 | 7 | 3 | 675972 | 2 | 431952 | 9 | -6186 | 9.159 | 5.405 | 108 | 33.0 | 0.0810 | 0.0392 | 15.7 | 6.4 |
| 1996 | 7 | 4 | 506670 | 2 | 330033 | 9 | -21911 | 8.730 | 4.921 | 79 | 27.2 | 0.0482 | 0.0479 | 8.7 | 5.8 |
| 1996 | 7 | 5 | 446974 | 2 | 318414 | 9 | 75 | 8.684 | 4.358 | 72 | 24.2 | 0.0501 | 0.0569 | 7.3 | 5.7 |
| 1996 | 7 | 6 | 342174 | 2 | 281994 | 9 | -29847 | 8.272 | 3.833 | 89 | 22.5 | 0.0663 | 0.0653 | 12.9 | 5.4 |
| 1996 | 7 | 7 | 302347 | 2 | 236494 | 9 | -22768 | 8.416 | 3.299 | 91 | 22.9 | 0.0672 | 0.0739 | 13.5 | 5.4 |
| 1996 | 7 | 8 | 293201 | 2 | 204026 | 9 | -2459 | 8.289 | 2.726 | 112 | 23.6 | 0.0985 | 0.0831 | 19.5 | 5.5 |
| 1996 | 7 | 9 | 264352 | 2 | 194124 | 9 | 1341 | 8.565 | 2.183 | 101 | 23.6 | 0.0702 | 0.0918 | 15.3 | 5.3 |
| 1996 | 7 | 10 | 390547 | 2 | 207159 | 9 | 22581 | 8.070 | 1.781 | 89 | 25.4 | 0.1927 | 0.0938 | 11.3 | 5.4 |
| 1996 | 7 | 11 | 750028 | 2 | 464193 | 9 | 47547 | 12.848 | 2.494 | 72 | 28.0 | 0.1279 | 0.0728 | 6.6 | 5.7 |
| 1996 | 7 | 12 | 604091 | 2 | 440483 | 9 | 13675 | 11.688 | 4.130 | 76 | 29.6 | 0.0623 | 0.0598 | 7.9 | 5.7 |
| 1997 | 8 | 1 | 400316 | 2 | 341331 | 8 | 1081 | 9.359 | 4.980 | 50 | 31.0 | 0.0716 | 0.0541 | 7.4 | 5.7 |
| 1997 | 8 | 2 | 674954 | 2 | 645134 | 8 | 35251 | 10.883 | 5.324 | 69 | 31.8 | 0.1057 | 0.0425 | 12.5 | 5.8 |
| 1997 | 8 | 3 | 672283 | 2 | 462355 | 8 | 3167 | 9.819 | 5.483 | 82 | 29.0 | 0.0573 | 0.0328 | 16.1 | 5.5 |
| 1997 | 8 | 4 | 494524 | 2 | 299514 | 8 | 3764 | 8.550 | 4.971 | 49 | 23.3 | 0.0554 | 0.0311 | 7.3 | 4.9 |
| 1997 | 8 | 5 | 578913 | 2 | 384087 | 8 | 9539 | 7.740 | 4.189 | 42 | 20.3 | 0.0634 | 0.0325 | 5.5 | 4.8 |
| 1997 | 8 | 6 | 413726 | 2 | 252534 | 8 | -16284 | 7.597 | 3.482 | 66 | 18.5 | 0.0725 | 0.0338 | 12.3 | 4.5 |
| 1997 | 8 | 7 | 352230 | 2 | 201256 | 8 | -18594 | 8.271 | 2.852 | 69 | 19.0 | 0.0812 | 0.0358 | 13.1 | 4.5 |
| 1997 | 8 | 8 | 222775 | 2 | 170682 | 8 | -23885 | 8.307 | 2.621 | 97 | 19.6 | 0.0714 | 0.0417 | 21.1 | 4.5 |
| 1997 | 8 | 9 | 233665 | 2 | 202412 | 8 | -1118 | 7.903 | 2.467 | 80 | 19.7 | 0.0777 | 0.0481 | 16.0 | 4.4 |
| 1997 | 8 | 10 | 335481 | 2 | 213536 | 8 | 29736 | 8.238 | 2.314 | 62 | 21.4 | 0.0994 | 0.0543 | 10.8 | 4.5 |
| 1997 | 8 | 11 | 330687 | 2 | 212062 | 8 | 14942 | 8.567 | 2.219 | 40 | 24.1 | 0.0806 | 0.0551 | 4.5 | 4.8 |
| 1997 | 8 | 12 | 436711 | 2 | 268618 | 8 | 24407 | 11.169 | 3.242 | 45 | 25.6 | 0.0762 | 0.0534 | 6.1 | 4.8 |
| 1998 | 9 | 1 | 734850 | 2 | 504964 | 8 | 28581 | 13.349 | 4.783 | 79 | 57.3 | 0.0818 | 0.0520 | 10.7 | 11.1 |
| 1998 | 9 | 2 | 683925 | 2 | 515762 | 8 | 23811 | 11.849 | 5.720 | 101 | 58.9 | 0.1217 | 0.0477 | 17.0 | 11.3 |
| 1998 | 9 | 3 | 930017 | 2 | 699945 | 8 | 17476 | 9.996 | 7.500 | 119 | 53.3 | 0.2144 | 0.0551 | 21.4 | 10.7 |
| 1998 | 9 | 4 | 710053 | 2 | 460371 | 8 | 19563 | 9.368 | 7.020 | 78 | 42.0 | 0.0934 | 0.0624 | 10.4 | 9.4 |
| 1998 | 9 | 5 | 478405 | 2 | 273519 | 8 | -22991 | 9.070 | 5.901 | 68 | 36.1 | 0.0817 | 0.0679 | 8.1 | 9.2 |
| 1998 | 9 | 6 | 342091 | 2 | 200246 | 8 | -7713 | 8.409 | 4.630 | 95 | 32.7 | 0.0691 | 0.0742 | 16.8 | 8.6 |
| 1998 | 9 | 7 | 339268 | 2 | 194372 | 8 | 9688 | 8.152 | 3.424 | 99 | 33.5 | 0.0735 | 0.0801 | 17.8 | 8.5 |
| 1998 | 9 | 8 | 338989 | 2 | 182689 | 8 | 1304 | 8.290 | 2.255 | 133 | 34.9 | 0.0644 | 0.0858 | 27.5 | 8.6 |
| 1998 | 9 | 9 | 377196 | 2 | 245080 | 8 | 15762 | 7.403 | 1.939 | 113 | 34.9 | 0.1313 | 0.0814 | 21.0 | 8.2 |
| 1998 | 9 | 10 | 976409 | 2 | 769075 | 8 | 71321 | 8.242 | 2.165 | 93 | 38.4 | 0.2085 | 0.0704 | 14.6 | 8.6 |
| 1998 | 9 | 11 | 1037113 | 2 | 666896 | 8 | 12781 | 9.437 | 3.788 | 66 | 43.6 | 0.1086 | 0.0748 | 7.0 | 9.1 |
| 1998 | 9 | 12 | 763451 | 2 | 520490 | 8 | 25376 | 9.713 | 5.289 | 73 | 46.6 | 0.0832 | 0.0728 | 9.0 | 9.2 |
| 1999 | 10 | 1 | 1061521 | 2 | 825716 | 8 | 33127 | 9.170 | 5.795 | 95 | 57.3 | 0.0888 | 0.0660 | 12.3 | 11.1 |
| 1999 | 10 | 2 | 762409 | 2 | 592226 | 8 | 16209 | 9.217 | 6.236 | 118 | 58.9 | 0.0905 | 0.0625 | 18.5 | 11.3 |
| 1999 | 10 | 3 | 1202061 | 2 | 916804 | 8 | 34803 | 9.256 | 6.182 | 136 | 53.3 | 0.0852 | 0.0615 | 22.9 | 10.7 |
| 1999 | 10 | 4 | 756658 | 2 | 516633 | 8 | -4173 | 8.210 | 5.762 | 93 | 42.0 | 0.0980 | 0.0654 | 11.9 | 9.4 |
| 1999 | 10 | 5 | 445460 | 2 | 297382 | 8 | -18743 | 8.428 | 5.333 | 83 | 36.1 | 0.1017 | 0.0694 | 9.7 | 9.2 |
| 1999 | 10 | 6 | 410340 | 2 | 291240 | 8 | 11477 | 7.210 | 4.899 | 110 | 32.6 | 0.0998 | 0.0735 | 18.4 | 8.6 |
| 1999 | 10 | 7 | 413590 | 2 | 299269 | 8 | -13936 | 5.186 | 4.499 | 113 | 33.5 | 0.0937 | 0.0772 | 19.3 | 8.5 |
| 1999 | 10 | 8 | 332181 | 2 | 248061 | 8 | -4173 | 7.749 | 4.031 | 147 | 34.9 | 0.0833 | 0.0816 | 28.8 | 8.6 |
| 1999 | 10 | 9 | 443551 | 2 | 314792 | 8 | 26568 | 7.761 | 3.591 | 129 | 34.9 | 0.0865 | 0.0857 | 22.3 | 8.2 |
| 1999 | 10 | 10 | 946701 | 2 | 733099 | 8 | 35623 | 7.830 | 3.228 | 109 | 38.4 | 0.0979 | 0.0891 | 16.0 | 8.6 |
| 1999 | 10 | 11 | 566849 | 2 | 433254 | 8 | 9875 | 8.072 | 3.194 | 82 | 43.6 | 0.1113 | 0.0810 | 8.5 | 9.1 |
| 1999 | 10 | 12 | 1403960 | 2 | 1171279 | 8 | 69868 | 8.680 | 4.139 | 88 | 46.6 | 0.1200 | 0.0771 | 10.6 | 9.2 |
| 2000 | 11 | 1 | 1098010 | 2 | 866954 | 8 | 23252 | 7.926 | 4.946 | 81 | 63.8 | 0.0893 | 0.0744 | 11.2 | 12.4 |
| 2000 | 11 | 2 | 1171588 | 2 | 823332 | 8 | 28990 | 7.857 | 5.220 | 110 | 65.5 | 0.0704 | 0.0552 | 18.2 | 12.5 |
| 2000 | 11 | 3 | 1077538 | 2 | 729430 | 8 | 17700 | 7.428 | 5.400 | 131 | 59.2 | 0.0629 | 0.0401 | 23.1 | 11.9 |
| 2000 | 11 | 4 | 641809 | 2 | 420086 | 8 | -4993 | 8.029 | 5.093 | 80 | 46.3 | 0.0686 | 0.0389 | 10.9 | 10.4 |
| 2000 | 11 | 5 | 522797 | 2 | 301732 | 8 | -16247 | 8.354 | 4.255 | 68 | 39.5 | 0.0694 | 0.0338 | 8.4 | 10.2 |
| 2000 | 11 | 6 | 497384 | 2 | 245669 | 8 | -13042 | 7.886 | 3.279 | 103 | 35.6 | 0.0697 | 0.0751 | 18.0 | 9.5 |
| 2000 | 11 | 7 | 428400 | 2 | 201766 | 8 | -10024 | 7.956 | 2.329 | 108 | 36.6 | 0.0652 | 0.1109 | 19.1 | 9.4 |
| 2000 | 11 | 8 | 386321 | 2 | 197379 | 8 | -5105 | 8.524 | 1.897 | 151 | 38.1 | 0.0698 | 0.0717 | 29.9 | 9.5 |
| 2000 | 11 | 9 | 442450 | 2 | 229191 | 8 | 15129 | 8.867 | 1.506 | 126 | 38.2 | 0.0762 | 0.0930 | 22.7 | 9.1 |
| 2000 | 11 | 10 | 476797 | 2 | 299343 | 8 | 27537 | 7.706 | 1.551 | 99 | 42.1 | 0.1035 | 0.0879 | 15.6 | 9.5 |
| 2000 | 11 | 11 | 828174 | 2 | 613556 | 8 | 34990 | 8.535 | 1.954 | 65 | 48.1 | 0.0946 | 0.0582 | 7.1 | 10.1 |
| 2000 | 11 | 12 | 927235 | 2 | 584848 | 8 | 31040 | 7.866 | 3.176 | 73 | 51.6 | 0.1115 | 0.0523 | 9.4 | 10.2 |
| 2001 | 12 | 1 | 787782 | 2 | 609434 | 8 | 15390 | 8.308 | 4.434 | 81 | 63.9 | 0.2144 | 0.0491 | 10.9 | 12.4 |
| 2001 | 12 | 2 | 782294 | 2 | 629608 | 8 | 21612 | 8.418 | 4.985 | 109 | 65.7 | 0.1145 | 0.0518 | 17.8 | 12.6 |
| 2001 | 12 | 3 | 669588 | 2 | 522386 | 8 | 9353 | 8.653 | 5.206 | 131 | 59.4 | 0.0731 | 0.0445 | 22.7 | 11.9 |
| 2001 | 12 | 4 | 657973 | 2 | 482767 | 8 | 11142 | 8.357 | 4.597 | 80 | 46.4 | 0.0767 | 0.0392 | 10.6 | 10.5 |
| 2001 | 12 | 5 | 442336 | 2 | 294619 | 8 | -23066 | 7.571 | 3.849 | 68 | 39.6 | 0.1161 | 0.0298 | 8.1 | 10.2 |
| 2001 | 12 | 6 | 390409 | 2 | 272609 | 8 | -5888 | 6.990 | 3.029 | 103 | 35.7 | 0.1044 | 0.0381 | 17.6 | 9.5 |
| 2001 | 12 | 7 | 326973 | 2 | 269613 | 8 | -22991 | 6.917 | 2.076 | 108 | 36.7 | 0.1115 | 0.0551 | 18.7 | 9.4 |
| 2001 | 12 | 8 | 366118 | 2 | 256225 | 8 | 3614 | 7.763 | 1.679 | 151 | 38.2 | 0.0876 | 0.0999 | 29.5 | 9.5 |
| 2001 | 12 | 9 | 590276 | 2 | 384705 | 8 | 45461 | 7.602 | 1.798 | 125 | 38.2 | 0.0958 | 0.1199 | 22.3 | 9.1 |
| 2001 | 12 | 10 | 736990 | 2 | 573988 | 8 | 20718 | 7.312 | 2.074 | 99 | 42.2 | 0.1315 | 0.0871 | 15.3 | 9.5 |
| 2001 | 12 | 11 | 649197 | 2 | 519725 | 8 | 22209 | 7.482 | 2.280 | 65 | 48.2 | 0.1232 | 0.0563 | 6.8 | 10.1 |
| 2001 | 12 | 12 | 694804 | 2 | 559615 | 8 | 32829 | 7.971 | 2.987 | 73 | 51.7 | 0.1024 | 0.0537 | 9.1 | 10.2 |
| 2002 | 13 | 1 | 1078804 | 2 | 930463 | 9 | 40057 | 7.760 | 4.656 | 90 | 57.3 | 0.0994 | 0.0598 | 13.7 | 11.1 |
| 2002 | 13 | 2 | 1612752 | 2 | 1273913 | 9 | 63198 | 7.410 | 5.255 | 122 | 58.9 | 0.0752 | 0.0580 | 21.7 | 11.3 |
| 2002 | 13 | 3 | 1795076 | 2 | 1282321 | 9 | 10285 | 6.159 | 4.944 | 147 | 53.3 | 0.0928 | 0.0622 | 27.3 | 10.7 |
| 2002 | 13 | 4 | 713548 | 2 | 534244 | 9 | -10024 | 7.459 | 4.518 | 88 | 42.0 | 0.0743 | 0.0471 | 13.3 | 9.4 |
| 2002 | 13 | 5 | 627609 | 2 | 352052 | 9 | -9539 | 8.166 | 4.161 | 75 | 36.1 | 0.0726 | 0.0340 | 10.5 | 9.2 |
| 2002 | 13 | 6 | 676556 | 2 | 342023 | 9 | 18818 | 5.857 | 3.375 | 115 | 32.6 | 0.1102 | 0.0384 | 21.5 | 8.6 |
| 2002 | 13 | 7 | 871268 | 2 | 696518 | 9 | 19526 | 5.910 | 2.567 | 120 | 33.5 | 0.1125 | 0.0692 | 22.7 | 8.5 |
| 2002 | 13 | 8 | 626794 | 2 | 386559 | 9 | -4807 | 6.398 | 2.005 | 170 | 34.9 | 0.1187 | 0.0551 | 35.0 | 8.6 |
| 2002 | 13 | 9 | 458379 | 2 | 246137 | 9 | -16582 | 7.511 | 1.736 | 141 | 34.9 | 0.1200 | 0.1200 | 26.7 | 8.2 |
| 2002 | 13 | 10 | 663787 | 2 | 444920 | 9 | 38642 | 8.224 | 1.860 | 110 | 38.4 | 0.1091 | 0.0692 | 18.7 | 8.6 |
| 2002 | 13 | 11 | 830415 | 2 | 603649 | 9 | 34319 | 8.464 | 2.682 | 71 | 43.6 | 0.0963 | 0.0508 | 9.0 | 9.1 |
| 2002 | 13 | 12 | 726688 | 2 | 538515 | 9 | 11365 | 8.363 | 3.655 | 81 | 46.6 | 0.0646 | 0.0449 | 11.6 | 9.2 |
| 2003 | 14 | 1 | 845535 | 2 | 613952 | 10 | 27351 | 8.876 | 4.433 | 86 | 16.2 | 0.0825 | 0.0402 | 12.4 | 3.6 |
| 2003 | 14 | 2 | 681970 | 2 | 477168 | 10 | 3167 | 9.514 | 4.503 | 110 | 16.3 | 0.0547 | 0.0352 | 18.2 | 3.6 |
| 2003 | 14 | 3 | 608393 | 2 | 447977 | 10 | -6148 | 7.625 | 4.677 | 129 | 15.8 | 0.0555 | 0.0270 | 22.2 | 3.5 |
| 2003 | 14 | 4 | 540259 | 2 | 369254 | 10 | -1751 | 7.249 | 4.210 | 85 | 14.8 | 0.0522 | 0.0192 | 12.0 | 3.4 |
| 2003 | 14 | 5 | 634356 | 2 | 406068 | 10 | 6037 | 7.054 | 3.633 | 75 | 14.2 | 0.0740 | 0.0176 | 9.9 | 3.3 |
| 2003 | 14 | 6 | 468594 | 2 | 263470 | 10 | -8272 | 6.551 | 3.075 | 103 | 13.9 | 0.0884 | 0.0242 | 18.1 | 3.3 |
| 2003 | 14 | 7 | 510073 | 2 | 296464 | 10 | -6745 | 7.614 | 2.561 | 107 | 14.0 | 0.0779 | 0.0518 | 19.0 | 3.2 |
| 2003 | 14 | 8 | 377194 | 2 | 220749 | 10 | -16768 | 13.194 | 1.998 | 142 | 14.1 | 0.0751 | 0.0780 | 27.7 | 3.3 |
| 2003 | 14 | 9 | 369657 | 2 | 192821 | 10 | -5478 | 11.885 | 1.564 | 122 | 14.1 | 0.0801 | 0.1032 | 21.6 | 3.2 |
| 2003 | 14 | 10 | 372860 | 2 | 177197 | 10 | 7825 | 8.558 | 1.129 | 101 | 14.5 | 0.0722 | 0.0649 | 15.8 | 3.3 |
| 2003 | 14 | 11 | 412644 | 2 | 221993 | 10 | 23476 | 8.618 | 1.352 | 73 | 14.9 | 0.0869 | 0.0662 | 8.9 | 3.3 |
| 2003 | 14 | 12 | 575831 | 2 | 353828 | 10 | 38977 | 9.059 | 2.073 | 80 | 15.2 | 0.0862 | 0.0708 | 10.9 | 3.3 |
| 2004 | 15 | 1 | 896204 | 2 | 574771 | 6 | 32195 | 10.074 | 3.220 | 93 | 18.0 | 0.0908 | 0.0793 | 14.1 | 3.6 |
| 2004 | 15 | 2 | 1028900 | 2 | 643255 | 6 | 24482 | 8.410 | 4.358 | 124 | 18.2 | 0.0894 | 0.0682 | 21.8 | 3.6 |
| 2004 | 15 | 3 | 847090 | 2 | 530020 | 6 | 15390 | 7.944 | 4.668 | 148 | 17.5 | 0.0908 | 0.0546 | 27.1 | 3.6 |
| 2004 | 15 | 4 | 649176 | 2 | 422317 | 6 | -2497 | 7.247 | 4.395 | 91 | 16.0 | 0.1127 | 0.0536 | 13.6 | 3.4 |
| 2004 | 15 | 5 | 498109 | 2 | 311167 | 6 | -11514 | 7.687 | 3.357 | 78 | 15.2 | 0.1045 | 0.0642 | 10.9 | 3.4 |
| 2004 | 15 | 6 | 414671 | 2 | 257963 | 6 | -2348 | 8.207 | 2.301 | 117 | 14.7 | 0.1026 | 0.1099 | 21.6 | 3.3 |
| 2004 | 15 | 7 | 402499 | 2 | 272684 | 6 | -1155 | 8.124 | 1.301 | 122 | 14.9 | 0.0911 | 0.0732 | 22.7 | 3.3 |
| 2004 | 15 | 8 | 394858 | 2 | 276497 | 6 | 5105 | 7.343 | 1.102 | 169 | 15.0 | 0.1184 | 0.1096 | 34.4 | 3.3 |
| 2004 | 15 | 9 | 486885 | 2 | 366933 | 6 | 14980 | 7.138 | 1.337 | 142 | 15.0 | 0.1114 | 0.1410 | 26.5 | 3.2 |
| 2004 | 15 | 10 | 840845 | 2 | 627344 | 6 | 41064 | 7.097 | 1.695 | 113 | 15.5 | 0.1169 | 0.1443 | 18.7 | 3.3 |
| 2004 | 15 | 11 | 971124 | 2 | 746922 | 6 | 25376 | 8.057 | 3.047 | 75 | 16.2 | 0.0878 | 0.1150 | 9.5 | 3.4 |
| 2004 | 15 | 12 | 903804 | 2 | 672684 | 6 | 34468 | 8.089 | 3.971 | 84 | 16.6 | 0.0839 | 0.0864 | 12.1 | 3.4 |
| 2005 | 16 | 1 | 1318629 | 2 | 1026846 | 9 | 35623 | 7.369 | 4.621 | 87 | 30.9 | 0.0789 | 0.0774 | 12.6 | 6.1 |
| 2005 | 16 | 2 | 766334 | 2 | 535151 | 9 | 24668 | 7.549 | 4.896 | 111 | 31.5 | 0.0724 | 0.0650 | 18.6 | 6.1 |
| 2005 | 16 | 3 | 1222094 | 2 | 734021 | 9 | 8272 | 6.797 | 4.707 | 131 | 29.2 | 0.1803 | 0.0513 | 22.7 | 5.9 |
| 2005 | 16 | 4 | 762771 | 2 | 425045 | 9 | -11030 | 7.061 | 4.235 | 85 | 24.5 | 0.0610 | 0.0363 | 12.2 | 5.4 |
| 2005 | 16 | 5 | 558106 | 2 | 336469 | 9 | -11477 | 7.207 | 3.551 | 75 | 22.1 | 0.0504 | 0.0252 | 10.0 | 5.3 |
| 2005 | 16 | 6 | 478510 | 2 | 259499 | 9 | -15986 | 7.341 | 2.788 | 104 | 20.6 | 0.0718 | 0.0327 | 18.5 | 5.0 |
| 2005 | 16 | 7 | 401677 | 2 | 293454 | 9 | -4956 | 6.593 | 2.017 | 108 | 21.0 | 0.0786 | 0.0725 | 19.4 | 5.0 |
| 2005 | 16 | 8 | 442761 | 2 | 252738 | 9 | -4062 | 7.455 | 1.691 | 145 | 21.5 | 0.0728 | 0.1138 | 28.4 | 5.0 |
| 2005 | 16 | 9 | 298689 | 2 | 168284 | 9 | -14234 | 7.817 | 1.356 | 124 | 21.6 | 0.0774 | 0.1380 | 22.1 | 4.9 |
| 2005 | 16 | 10 | 451918 | 2 | 220619 | 9 | 15203 | 7.807 | 1.406 | 102 | 23.0 | 0.0828 | 0.1241 | 16.1 | 5.0 |
| 2005 | 16 | 11 | 552322 | 2 | 342288 | 9 | 29736 | 7.533 | 1.683 | 73 | 25.2 | 0.1064 | 0.0823 | 9.0 | 5.2 |
| 2005 | 16 | 12 | 570640 | 2 | 378673 | 9 | 20271 | 7.735 | 2.384 | 80 | 26.4 | 0.1036 | 0.0651 | 11.0 | 5.3 |
| 2006 | 17 | 1 | 547980 | 2 | 367250 | 9 | 17961 | 7.483 | 2.751 | 96 | 16.6 | 0.0931 | 0.0527 | 14.2 | 3.3 |
| 2006 | 17 | 2 | 885036 | 2 | 552038 | 9 | 21873 | 7.609 | 3.453 | 127 | 16.7 | 0.0857 | 0.0436 | 21.8 | 3.3 |
| 2006 | 17 | 3 | 629744 | 2 | 368065 | 9 | 9651 | 7.808 | 4.313 | 151 | 16.2 | 0.0834 | 0.0467 | 27.1 | 3.3 |
| 2006 | 17 | 4 | 710755 | 2 | 431291 | 9 | 11551 | 6.925 | 4.604 | 94 | 15.0 | 0.1039 | 0.0394 | 13.8 | 3.2 |
| 2006 | 17 | 5 | 601117 | 2 | 472552 | 9 | 5478 | 5.147 | 4.010 | 81 | 14.4 | 0.1314 | 0.0369 | 11.1 | 3.1 |
| 2006 | 17 | 6 | 387824 | 2 | 321610 | 9 | -27947 | 6.240 | 3.460 | 119 | 14.1 | 0.1523 | 0.0498 | 21.7 | 3.1 |
| 2006 | 17 | 7 | 265365 | 2 | 227840 | 9 | -36443 | 6.001 | 2.542 | 124 | 14.2 | 0.1998 | 0.0679 | 22.8 | 3.1 |
| 2006 | 17 | 8 | 355999 | 2 | 209977 | 9 | 19190 | 6.390 | 2.003 | 171 | 14.3 | 0.1659 | 0.1985 | 34.4 | 3.1 |
| 2006 | 17 | 9 | 279854 | 2 | 140299 | 9 | -14980 | 5.014 | 1.864 | 144 | 14.3 | 0.0663 | 0.1360 | 26.5 | 3.0 |
| 2006 | 17 | 10 | 412574 | 2 | 260004 | 9 | 40169 | 5.741 | 1.822 | 115 | 14.7 | 0.1090 | 0.0921 | 18.8 | 3.1 |
| 2006 | 17 | 11 | 685790 | 2 | 463040 | 9 | 34394 | 8.602 | 2.383 | 78 | 15.2 | 0.1079 | 0.0725 | 9.7 | 3.1 |
| 2006 | 17 | 12 | 1204315 | 2 | 1061149 | 9 | 54329 | 6.824 | 3.322 | 87 | 15.5 | 0.1835 | 0.0768 | 12.2 | 3.1 |
| 2007 | 18 | 1 | 1435423 | 2 | 1358991 | 8 | 71246 | 6.473 | 4.371 | 101 | 14.0 | 0.1198 | 0.0829 | 14.8 | 2.8 |
| 2007 | 18 | 2 | 1017285 | 2 | 861289 | 8 | 42964 | 6.211 | 4.566 | 134 | 14.1 | 0.1345 | 0.0728 | 22.9 | 2.8 |
| 2007 | 18 | 3 | 1494812 | 2 | 1123827 | 8 | 8347 | 5.964 | 4.361 | 159 | 13.8 | 0.1213 | 0.0572 | 28.5 | 2.8 |
| 2007 | 18 | 4 | 631771 | 2 | 463898 | 8 | -24221 | 6.973 | 3.984 | 99 | 13.3 | 0.1121 | 0.0282 | 14.4 | 2.8 |
| 2007 | 18 | 5 | 406755 | 2 | 291606 | 8 | -6260 | 7.903 | 3.311 | 85 | 13.1 | 0.1085 | 0.0338 | 11.5 | 2.8 |
| 2007 | 18 | 6 | 442002 | 2 | 306837 | 8 | -3093 | 7.881 | 2.458 | 126 | 12.9 | 0.0955 | 0.0537 | 22.7 | 2.7 |
| 2007 | 18 | 7 | 641624 | 2 | 417951 | 8 | 8757 | 6.526 | 1.605 | 131 | 13.0 | 0.0843 | 0.0767 | 23.9 | 2.7 |
| 2007 | 18 | 8 | 444206 | 2 | 250589 | 8 | -11589 | 7.224 | 0.849 | 180 | 13.0 | 0.0712 | 0.0929 | 36.2 | 2.7 |
| 2007 | 18 | 9 | 449375 | 2 | 281674 | 8 | 13862 | 7.132 | 0.935 | 152 | 13.0 | 0.0793 | 0.1369 | 27.9 | 2.7 |
| 2007 | 18 | 10 | 515405 | 2 | 374505 | 8 | 3056 | 7.028 | 1.041 | 121 | 13.2 | 0.1001 | 0.1039 | 19.7 | 2.7 |
| 2007 | 18 | 11 | 524090 | 2 | 394190 | 8 | 16768 | 7.210 | 1.417 | 82 | 13.4 | 0.0767 | 0.0500 | 10.0 | 2.7 |
| 2007 | 18 | 12 | 848381 | 2 | 691831 | 8 | 37188 | 7.562 | 2.213 | 92 | 13.5 | 0.0935 | 0.0509 | 12.7 | 2.8 |
| 2008 | 19 | 1 | 1196100 | 2 | 906197 | 6 | 44194 | 6.633 | 3.190 | 96 | 14.1 | 0.0936 | 0.0753 | 12.0 | 3.7 |
| 2008 | 19 | 2 | 1010150 | 2 | 736423 | 6 | 20308 | 6.141 | 4.028 | 126 | 14.1 | 0.0898 | 0.0644 | 19.3 | 3.8 |
| 2008 | 19 | 3 | 1233842 | 2 | 842395 | 6 | 39722 | 5.843 | 3.559 | 149 | 13.9 | 0.0747 | 0.0484 | 24.5 | 3.7 |
| 2008 | 19 | 4 | 897055 | 2 | 539340 | 6 | -484 | 5.519 | 3.253 | 95 | 13.4 | 0.0734 | 0.0297 | 11.7 | 3.4 |
| 2008 | 19 | 5 | 511681 | 2 | 320505 | 6 | -39946 | 6.661 | 2.825 | 82 | 13.1 | 0.0809 | 0.0407 | 9.1 | 3.4 |
| 2008 | 19 | 6 | 345842 | 2 | 220781 | 6 | -32270 | 7.019 | 1.736 | 119 | 13.0 | 0.0602 | 0.0617 | 19.2 | 3.3 |
| 2008 | 19 | 7 | 391709 | 2 | 286940 | 6 | -20532 | 7.198 | 0.972 | 123 | 13.0 | 0.0454 | 0.0912 | 20.3 | 3.3 |
| 2008 | 19 | 8 | 527535 | 2 | 373817 | 6 | 35474 | 6.022 | 0.712 | 168 | 13.1 | 0.0741 | 0.2532 | 31.6 | 3.3 |
| 2008 | 19 | 9 | 488202 | 2 | 420716 | 6 | 3391 | 6.139 | 0.886 | 143 | 13.1 | 0.0774 | 0.2576 | 24.0 | 3.2 |
| 2008 | 19 | 10 | 531934 | 2 | 589130 | 6 | 33015 | 6.428 | 1.046 | 115 | 13.2 | 0.0851 | 0.0912 | 16.6 | 3.3 |
| 2008 | 19 | 11 | 846300 | 2 | 768182 | 6 | 28730 | 7.076 | 1.926 | 79 | 13.5 | 0.0884 | 0.0634 | 7.7 | 3.4 |
| 2008 | 19 | 12 | 663513 | 2 | 454224 | 6 | 8272 | 7.596 | 2.907 | 88 | 13.6 | 0.0794 | 0.0555 | 10.1 | 3.4 |
| 2009 | 20 | 1 | 559198 | 2 | 435228 | 6 | 15017 | 7.542 | 3.653 | 89 | 17.9 | 0.0871 | 0.0510 | 11.7 | 2.9 |
| 2009 | 20 | 2 | 410219 | 2 | 352457 | 6 | 8570 | 7.761 | 4.136 | 116 | 18.1 | 0.0654 | 0.0443 | 18.5 | 2.9 |
| 2009 | 20 | 3 | 498758 | 2 | 397533 | 6 | 10061 | 8.234 | 3.698 | 138 | 17.4 | 0.0640 | 0.0413 | 23.3 | 2.8 |
| 2009 | 20 | 4 | 385354 | 2 | 253798 | 6 | -26568 | 8.536 | 3.609 | 87 | 15.9 | 0.0558 | 0.0251 | 11.4 | 2.7 |
| 2009 | 20 | 5 | 390941 | 2 | 255096 | 6 | -10508 | 7.472 | 2.962 | 75 | 15.2 | 0.0626 | 0.0211 | 9.0 | 2.7 |
| 2009 | 20 | 6 | 323610 | 2 | 207080 | 6 | -28804 | 7.569 | 1.868 | 109 | 14.7 | 0.0540 | 0.0349 | 18.3 | 2.7 |
| 2009 | 20 | 7 | 299064 | 2 | 248413 | 6 | 1826 | 6.857 | 1.297 | 114 | 14.8 | 0.0526 | 0.0654 | 19.4 | 2.6 |
| 2009 | 20 | 8 | 357387 | 2 | 218032 | 6 | -783 | 6.487 | 1.395 | 156 | 15.0 | 0.0511 | 0.0795 | 29.8 | 2.7 |
| 2009 | 20 | 9 | 384116 | 2 | 241653 | 6 | 2459 | 6.227 | 1.434 | 132 | 15.0 | 0.0654 | 0.1695 | 22.8 | 2.6 |
| 2009 | 20 | 10 | 385499 | 2 | 350163 | 6 | 26792 | 6.681 | 1.361 | 106 | 15.5 | 0.0625 | 0.1104 | 15.9 | 2.6 |
| 2009 | 20 | 11 | 844647 | 2 | 539846 | 6 | 57273 | 7.322 | 2.036 | 73 | 16.1 | 0.0925 | 0.0646 | 7.7 | 2.7 |
| 2009 | 20 | 12 | 1082344 | 2 | 758553 | 6 | 31263 | 7.373 | 3.249 | 81 | 16.5 | 0.0834 | 0.0630 | 9.9 | 2.7 |
| 2010 | 21 | 1 | 600434 | 2 | 413476 | 7 | 5738 | 7.396 | 4.250 | 83 | 13.9 | 0.0909 | 0.0656 | 11.9 | 2.3 |
| 2010 | 21 | 2 | 366441 | 2 | 275532 | 7 | 10098 | 8.241 | 4.612 | 110 | 13.9 | 0.1080 | 0.0584 | 18.6 | 2.3 |
| 2010 | 21 | 3 | 829653 | 2 | 654816 | 7 | 5515 | 7.058 | 4.594 | 131 | 13.7 | 0.0879 | 0.0504 | 23.3 | 2.3 |
| 2010 | 21 | 4 | 476588 | 2 | 363367 | 7 | -11701 | 7.159 | 4.084 | 81 | 13.2 | 0.0672 | 0.0263 | 11.5 | 2.3 |
| 2010 | 21 | 5 | 400905 | 2 | 309781 | 7 | -2720 | 7.617 | 3.131 | 70 | 13.0 | 0.0840 | 0.0291 | 9.1 | 2.3 |
| 2010 | 21 | 6 | 363504 | 2 | 294914 | 7 | -10359 | 6.503 | 2.501 | 104 | 12.9 | 0.0920 | 0.0427 | 18.4 | 2.2 |
| 2010 | 21 | 7 | 305255 | 2 | 228894 | 7 | -4919 | 7.452 | 1.672 | 108 | 12.9 | 0.0939 | 0.0726 | 19.4 | 2.2 |
| 2010 | 21 | 8 | 512212 | 2 | 454790 | 7 | 35102 | 6.416 | 1.521 | 150 | 13.0 | 0.1243 | 0.1185 | 29.7 | 2.2 |
| 2010 | 21 | 9 | 539071 | 2 | 436257 | 7 | 9018 | 5.869 | 1.895 | 125 | 13.0 | 0.1060 | 0.1153 | 22.8 | 2.2 |
| 2010 | 21 | 10 | 510574 | 2 | 459112 | 7 | 27053 | 6.801 | 1.830 | 100 | 13.1 | 0.0929 | 0.0681 | 16.0 | 2.2 |
| 2010 | 21 | 11 | 842472 | 2 | 676343 | 7 | 43970 | 7.188 | 2.450 | 67 | 13.3 | 0.0841 | 0.0523 | 7.9 | 2.3 |
| 2010 | 21 | 12 | 699684 | 2 | 554691 | 7 | 15091 | 6.796 | 3.528 | 75 | 13.4 | 0.0836 | 0.0552 | 10.1 | 2.3 |
| 2011 | 22 | 1 | 1100376 | 2 | 886668 | 6 | 18072 | 5.201 | 3.966 | 79 | 15.7 | 0.1115 | 0.0919 | 12.5 | 1.8 |
| 2011 | 22 | 2 | 844224 | 2 | 615809 | 6 | 16358 | 6.233 | 3.956 | 108 | 15.8 | 0.0808 | 0.0743 | 19.7 | 1.8 |
| 2011 | 22 | 3 | 555731 | 2 | 425559 | 6 | 410 | 6.897 | 3.821 | 130 | 15.3 | 0.0729 | 0.0417 | 24.7 | 1.8 |
| 2011 | 22 | 4 | 454780 | 2 | 298132 | 6 | -22171 | 6.886 | 3.430 | 77 | 14.4 | 0.0722 | 0.0300 | 12.1 | 1.8 |
| 2011 | 22 | 5 | 377627 | 2 | 231239 | 6 | -9688 | 6.398 | 2.630 | 65 | 14.0 | 0.0890 | 0.0377 | 9.5 | 1.8 |
| 2011 | 22 | 6 | 356206 | 2 | 203549 | 6 | -12148 | 7.414 | 1.878 | 102 | 13.7 | 0.0743 | 0.0371 | 19.5 | 1.8 |
| 2011 | 22 | 7 | 371292 | 2 | 217219 | 6 | 4472 | 7.352 | 1.342 | 107 | 13.7 | 0.0881 | 0.0758 | 20.6 | 1.8 |
| 2011 | 22 | 8 | 441605 | 2 | 311067 | 6 | 26419 | 6.175 | 1.263 | 151 | 13.9 | 0.1310 | 0.1496 | 31.6 | 1.8 |
| 2011 | 22 | 9 | 686631 | 2 | 471943 | 6 | 25935 | 5.824 | 1.477 | 125 | 13.9 | 0.1175 | 0.1460 | 24.2 | 1.8 |
| 2011 | 22 | 10 | 652059 | 2 | 528961 | 6 | 13042 | 6.280 | 1.619 | 97 | 14.1 | 0.1030 | 0.0882 | 16.9 | 1.8 |
| 2011 | 22 | 11 | 436275 | 2 | 283524 | 6 | 6894 | 7.085 | 1.895 | 62 | 14.6 | 0.1080 | 0.0701 | 8.2 | 1.8 |
| 2011 | 22 | 12 | 893581 | 2 | 704421 | 6 | 41734 | 6.390 | 2.391 | 71 | 14.8 | 0.0804 | 0.0664 | 10.6 | 1.8 |
| 2012 | 23 | 1 | 1143031 | 2 | 861433 | 8 | 38604 | 6.043 | 3.334 | 38 | 15.7 | 0.0699 | 0.0624 | 7.8 | 1.8 |
| 2012 | 23 | 2 | 682766 | 2 | 537528 | 8 | 8496 | 7.111 | 3.922 | 69 | 15.8 | 0.0621 | 0.0561 | 15.5 | 1.8 |
| 2012 | 23 | 3 | 638093 | 2 | 472121 | 8 | 0 | 6.425 | 3.711 | 90 | 15.4 | 0.0676 | 0.0395 | 20.8 | 1.8 |
| 2012 | 23 | 4 | 589209 | 2 | 346688 | 8 | 8459 | 6.444 | 3.448 | 38 | 14.5 | 0.0662 | 0.0248 | 7.7 | 1.8 |
| 2012 | 23 | 5 | 458337 | 2 | 266324 | 8 | -25152 | 6.385 | 2.903 | 27 | 14.0 | 0.0582 | 0.0353 | 5.0 | 1.8 |
| 2012 | 23 | 6 | 412864 | 2 | 253137 | 8 | 12036 | 7.118 | 2.350 | 67 | 13.7 | 0.0596 | 0.0344 | 15.1 | 1.8 |
| 2012 | 23 | 7 | 435738 | 2 | 264383 | 8 | -6707 | 6.961 | 2.337 | 72 | 13.8 | 0.0716 | 0.0313 | 16.3 | 1.8 |
| 2012 | 23 | 8 | 373732 | 2 | 244231 | 8 | 5068 | 7.076 | 1.690 | 121 | 13.9 | 0.0618 | 0.0911 | 28.5 | 1.8 |
| 2012 | 23 | 9 | 540867 | 2 | 384521 | 8 | 29810 | 6.542 | 1.981 | 90 | 13.9 | 0.0754 | 0.1337 | 20.8 | 1.8 |
| 2012 | 23 | 10 | 872268 | 2 | 628063 | 8 | 32829 | 6.068 | 2.105 | 59 | 14.2 | 0.0874 | 0.0968 | 13.0 | 1.8 |
| 2012 | 23 | 11 | 750016 | 2 | 504297 | 8 | 22395 | 6.476 | 2.653 | 21 | 14.6 | 0.0851 | 0.0705 | 3.5 | 1.8 |
| 2012 | 23 | 12 | 856323 | 2 | 595771 | 8 | 38381 | 6.945 | 3.323 | 30 | 14.8 | 0.0786 | 0.0579 | 5.8 | 1.8 |
| 2013 | 24 | 1 | 1011204 | 2 | 679189 | 8 | 22171 | 6.196 | 3.489 | 32 | 14.8 | 0.0693 | 0.0560 | 6.3 | 1.8 |
| 2013 | 24 | 2 | 728496 | 2 | 470192 | 8 | 5850 | 6.468 | 3.855 | 56 | 14.9 | 0.0559 | 0.0468 | 12.3 | 1.8 |
| 2013 | 24 | 3 | 506724 | 2 | 294307 | 8 | -7005 | 6.864 | 4.289 | 73 | 14.5 | 0.0612 | 0.0317 | 16.5 | 1.8 |
| 2013 | 24 | 4 | 437189 | 2 | 246609 | 8 | -11887 | 7.448 | 4.522 | 32 | 13.9 | 0.0619 | 0.0264 | 6.2 | 1.8 |
| 2013 | 24 | 5 | 439451 | 2 | 263358 | 8 | -5850 | 7.502 | 3.656 | 23 | 13.5 | 0.0363 | 0.0215 | 4.1 | 1.8 |
| 2013 | 24 | 6 | 359645 | 2 | 202856 | 8 | -16247 | 7.372 | 3.137 | 55 | 13.3 | 0.0379 | 0.0221 | 12.0 | 1.8 |
| 2013 | 24 | 7 | 317684 | 2 | 158225 | 8 | -37188 | 7.142 | 2.819 | 59 | 13.3 | 0.0480 | 0.0217 | 13.0 | 1.8 |
| 2013 | 24 | 8 | 327723 | 2 | 153822 | 8 | -13340 | 6.234 | 1.948 | 97 | 13.4 | 0.0546 | 0.0377 | 22.4 | 1.8 |
| 2013 | 24 | 9 | 330619 | 2 | 181860 | 8 | 11253 | 5.542 | 1.689 | 73 | 13.4 | 0.0544 | 0.0341 | 16.4 | 1.8 |
| 2013 | 24 | 10 | 359745 | 2 | 210890 | 8 | 33276 | 6.863 | 1.660 | 48 | 13.6 | 0.0603 | 0.0472 | 10.4 | 1.8 |
| 2013 | 24 | 11 | 495450 | 2 | 290670 | 8 | 17998 | 6.674 | 1.940 | 18 | 14.0 | 0.0775 | 0.0562 | 2.9 | 1.8 |
| 2013 | 24 | 12 | 748990 | 2 | 614876 | 8 | 46131 | 6.217 | 2.971 | 26 | 14.1 | 0.0794 | 0.0528 | 4.7 | 1.8 |
| 2014 | 25 | 1 | 1149954 | 2 | 778668 | 9 | 41250 | 6.040 | 3.849 | 34 | 20.5 | 0.0740 | 0.0542 | 5.5 | 3.2 |
| 2014 | 25 | 2 | 791034 | 2 | 539920 | 9 | 18371 | 5.669 | 3.813 | 60 | 27.1 | 0.0669 | 0.0478 | 10.6 | 4.7 |
| 2014 | 25 | 3 | 554569 | 2 | 383184 | 9 | -1379 | 5.814 | 3.590 | 78 | 31.5 | 0.0627 | 0.0279 | 14.1 | 5.8 |
| 2014 | 25 | 4 | 404106 | 2 | 252715 | 9 | -8570 | 6.424 | 3.354 | 34 | 19.6 | 0.0611 | 0.0239 | 5.4 | 3.2 |
| 2014 | 25 | 5 | 408634 | 2 | 262517 | 9 | 1491 | 7.001 | 3.095 | 24 | 16.9 | 0.0576 | 0.0245 | 3.6 | 2.6 |
| 2014 | 25 | 6 | 321230 | 2 | 198557 | 9 | -24407 | 6.536 | 2.225 | 59 | 25.4 | 0.0737 | 0.0355 | 10.3 | 4.7 |
| 2014 | 25 | 7 | 339387 | 2 | 184762 | 9 | -10955 | 6.312 | 1.870 | 63 | 26.5 | 0.0591 | 0.0346 | 11.1 | 4.9 |
| 2014 | 25 | 8 | 316760 | 2 | 205760 | 9 | 17588 | 6.178 | 1.509 | 104 | 37.0 | 0.0559 | 0.0487 | 19.2 | 7.4 |
| 2014 | 25 | 9 | 296803 | 2 | 183052 | 9 | 820 | 6.528 | 1.758 | 78 | 30.4 | 0.0551 | 0.0635 | 14.1 | 5.8 |
| 2014 | 25 | 10 | 435948 | 2 | 265587 | 9 | 34915 | 6.278 | 1.749 | 52 | 23.9 | 0.0721 | 0.0612 | 8.9 | 4.2 |
| 2014 | 25 | 11 | 495247 | 2 | 366578 | 9 | 19190 | 6.062 | 1.908 | 19 | 16.0 | 0.0751 | 0.0431 | 2.6 | 2.3 |
| 2014 | 25 | 12 | 882770 | 2 | 763368 | 9 | 48293 | 6.504 | 2.846 | 27 | 18.2 | 0.0697 | 0.0521 | 4.1 | 2.8 |
| 2015 | 26 | 1 | 1328951 | 2 | 1093526 | 8 | 45274 | 5.846 | 3.849 | 37 | 23.5 | 0.0705 | 0.0560 | 6.0 | 3.7 |
| 2015 | 26 | 2 | 715058 | 2 | 578116 | 8 | 6633 | 6.201 | 4.172 | 66 | 32.2 | 0.0639 | 0.0459 | 11.7 | 5.7 |
| 2015 | 26 | 3 | 686181 | 2 | 509387 | 8 | 15464 | 6.340 | 3.982 | 86 | 37.9 | 0.0576 | 0.0319 | 15.7 | 7.1 |
| 2015 | 26 | 4 | 571386 | 2 | 407743 | 8 | -15129 | 7.018 | 3.634 | 36 | 22.1 | 0.0792 | 0.0275 | 6.0 | 3.6 |
| 2015 | 26 | 5 | 574664 | 2 | 429138 | 8 | 14197 | 8.388 | 3.035 | 26 | 18.6 | 0.1116 | 0.0288 | 3.9 | 2.9 |
| 2015 | 26 | 6 | 565095 | 2 | 362357 | 8 | -7378 | 6.214 | 2.552 | 65 | 29.8 | 0.0782 | 0.0329 | 11.5 | 5.6 |
| 2015 | 26 | 7 | 416539 | 2 | 257082 | 8 | -1118 | 6.402 | 1.765 | 69 | 31.2 | 0.0686 | 0.0556 | 12.4 | 5.9 |
| 2015 | 26 | 8 | 379747 | 2 | 214261 | 8 | -9427 | 6.244 | 1.217 | 116 | 45.1 | 0.0835 | 0.0950 | 21.4 | 9.2 |
| 2015 | 26 | 9 | 764304 | 2 | 395645 | 8 | 32046 | 5.430 | 1.274 | 86 | 36.3 | 0.1288 | 0.1040 | 15.7 | 7.1 |
| 2015 | 26 | 10 | 573083 | 2 | 343938 | 8 | 6372 | 5.827 | 1.429 | 57 | 27.9 | 0.0939 | 0.0564 | 9.9 | 5.0 |
| 2015 | 26 | 11 | 988295 | 2 | 591798 | 8 | 58093 | 6.349 | 1.999 | 20 | 17.5 | 0.0728 | 0.0488 | 2.8 | 2.5 |
| 2015 | 26 | 12 | 1795026 | 2 | 1067415 | 8 | 57906 | 5.285 | 3.509 | 29 | 20.3 | 0.0668 | 0.0557 | 4.5 | 3.1 |
| 2016 | 27 | 1 | 1256295 | 2 | 798510 | 9 | 25749 | 5.098 | 3.564 | 39 | 21.5 | 0.0817 | 0.0661 | 7.1 | 3.1 |
| 2016 | 27 | 2 | 1138268 | 2 | 774631 | 9 | 20495 | 5.621 | 3.948 | 70 | 27.8 | 0.0614 | 0.0698 | 13.9 | 4.6 |
| 2016 | 27 | 3 | 839356 | 2 | 582515 | 9 | 4136 | 6.187 | 3.630 | 92 | 31.7 | 0.0511 | 0.0360 | 18.7 | 5.6 |
| 2016 | 27 | 4 | 975369 | 2 | 581753 | 9 | 25786 | 5.576 | 3.187 | 38 | 20.1 | 0.0562 | 0.0338 | 7.0 | 3.1 |
| 2016 | 27 | 5 | 824236 | 2 | 537426 | 9 | -19265 | 5.796 | 3.167 | 27 | 17.4 | 0.0681 | 0.0407 | 4.5 | 2.6 |
| 2016 | 27 | 6 | 603188 | 2 | 436572 | 9 | -8272 | 5.825 | 2.929 | 68 | 25.3 | 0.0784 | 0.0327 | 13.6 | 4.5 |
| 2016 | 27 | 7 | 727975 | 2 | 522077 | 9 | 6633 | 5.963 | 2.671 | 73 | 26.4 | 0.0842 | 0.0280 | 14.7 | 4.7 |
| 2016 | 27 | 8 | 489635 | 2 | 282414 | 9 | -8906 | 6.851 | 2.147 | 123 | 36.4 | 0.0730 | 0.0402 | 25.5 | 7.1 |
| 2016 | 27 | 9 | 428716 | 2 | 259463 | 9 | -11067 | 7.230 | 1.898 | 91 | 30.1 | 0.0690 | 0.0509 | 18.6 | 5.6 |
| 2016 | 27 | 10 | 606060 | 2 | 464552 | 9 | 28841 | 6.377 | 1.956 | 60 | 24.1 | 0.0777 | 0.0414 | 11.7 | 4.1 |
| 2016 | 27 | 11 | 861782 | 2 | 596946 | 9 | 36070 | 7.162 | 2.592 | 21 | 16.8 | 0.0651 | 0.0503 | 3.2 | 2.3 |
| 2016 | 27 | 12 | 824449 | 2 | 509805 | 9 | 20047 | 7.187 | 3.413 | 30 | 18.9 | 0.0575 | 0.0441 | 5.3 | 2.7 |
| 2017 | 28 | 1 | 804947 | 2 | 491858 | 7 | 16656 | 7.744 | 4.099 | 44 | 20.9 | 0.0612 | 0.0458 | 8.2 | 3.5 |
| 2017 | 28 | 2 | 635740 | 2 | 459718 | 7 | 25227 | 6.848 | 4.347 | 80 | 28.8 | 0.1191 | 0.0380 | 16.2 | 5.3 |
| 2017 | 28 | 3 | 876199 | 2 | 551942 | 7 | 9577 | 6.786 | 4.069 | 106 | 34.1 | 0.0584 | 0.0324 | 21.9 | 6.6 |
| 2017 | 28 | 4 | 628431 | 2 | 401014 | 7 | 2608 | 6.037 | 3.693 | 43 | 20.2 | 0.0562 | 0.0232 | 8.1 | 3.4 |
| 2017 | 28 | 5 | 508857 | 2 | 298062 | 7 | -27425 | 5.822 | 3.274 | 30 | 17.2 | 0.0637 | 0.0339 | 5.2 | 2.8 |
| 2017 | 28 | 6 | 496686 | 2 | 285380 | 7 | 7266 | 6.260 | 2.683 | 78 | 27.5 | 0.0700 | 0.0471 | 15.9 | 5.2 |
| 2017 | 28 | 7 | 445988 | 2 | 279822 | 7 | -2832 | 5.780 | 1.555 | 84 | 28.8 | 0.0906 | 0.0511 | 17.2 | 5.5 |
| 2017 | 28 | 8 | 530279 | 2 | 322185 | 7 | 9241 | 5.579 | 1.429 | 142 | 41.3 | 0.0675 | 0.0772 | 29.9 | 8.5 |
| 2017 | 28 | 9 | 658246 | 2 | 378977 | 7 | 21873 | 5.800 | 1.377 | 105 | 33.4 | 0.0550 | 0.0625 | 21.8 | 6.6 |
| 2017 | 28 | 10 | 892361 | 2 | 515848 | 7 | 37747 | 6.276 | 1.592 | 68 | 25.5 | 0.0592 | 0.0422 | 13.7 | 4.7 |
| 2017 | 28 | 11 | 721219 | 2 | 443691 | 7 | 26680 | 6.927 | 2.202 | 23 | 15.9 | 0.0684 | 0.0423 | 3.6 | 2.4 |
| 2017 | 28 | 12 | 797873 | 2 | 525417 | 7 | 26755 | 7.112 | 3.090 | 34 | 18.4 | 0.0516 | 0.0353 | 6.1 | 3.0 |
| 2018 | 29 | 1 | 1084382 | 2 | 730538 | 7 | 36927 | 6.228 | 3.758 | 32 | 17.6 | 0.0874 | 0.0352 | 4.6 | 2.5 |
| 2018 | 29 | 2 | 860261 | 2 | 629883 | 7 | 14682 | 6.566 | 4.110 | 55 | 22.9 | 0.0627 | 0.0330 | 8.8 | 3.2 |
| 2018 | 29 | 3 | 831398 | 2 | 534812 | 7 | 6633 | 6.805 | 3.894 | 72 | 26.6 | 0.0477 | 0.0267 | 11.7 | 3.8 |
| 2018 | 29 | 4 | 600998 | 2 | 330880 | 7 | -2422 | 6.830 | 3.736 | 31 | 17.3 | 0.0397 | 0.0258 | 4.6 | 2.5 |
| 2018 | 29 | 5 | 539290 | 2 | 299996 | 7 | -44305 | 6.196 | 3.317 | 23 | 15.3 | 0.0562 | 0.0238 | 3.1 | 2.2 |
| 2018 | 29 | 6 | 357580 | 2 | 216402 | 7 | -31711 | 6.746 | 2.748 | 54 | 22.3 | 0.0463 | 0.0347 | 8.6 | 3.2 |
| 2018 | 29 | 7 | 314021 | 2 | 165095 | 7 | -44082 | 5.744 | 1.722 | 58 | 23.1 | 0.0418 | 0.0540 | 9.3 | 3.3 |
| 2018 | 29 | 8 | 460415 | 2 | 234511 | 7 | 8906 | 4.338 | 1.707 | 95 | 31.6 | 0.0543 | 0.0716 | 15.8 | 4.5 |
| 2018 | 29 | 9 | 432283 | 2 | 247968 | 7 | 20457 | 5.992 | 1.714 | 72 | 26.2 | 0.0492 | 0.0593 | 11.6 | 3.8 |
| 2018 | 29 | 10 | 400042 | 2 | 200146 | 7 | 8682 | 6.868 | 1.502 | 48 | 20.9 | 0.0378 | 0.0308 | 7.4 | 3.0 |
| 2018 | 29 | 11 | 433270 | 2 | 202053 | 7 | 11104 | 7.121 | 1.678 | 18 | 14.4 | 0.0531 | 0.0399 | 2.3 | 2.0 |
| 2018 | 29 | 12 | 645584 | 2 | 365078 | 7 | 33425 | 8.997 | 2.492 | 25 | 16.0 | 0.0521 | 0.0410 | 3.5 | 2.3 |
| 2019 | 30 | 1 | 705730 | 2 | 428107 | 7 | 24109 | 8.512 | 3.582 | 52 | 13.1 | 0.0539 | 0.0397 | 6.3 | 1.9 |
| 2019 | 30 | 2 | 849209 | 2 | 502539 | 7 | 18072 | 7.961 | 4.149 | 98 | 15.8 | 0.0812 | 0.0510 | 12.4 | 2.2 |
| 2019 | 30 | 3 | 1235838 | 2 | 706376 | 7 | 38791 | 7.131 | 5.016 | 130 | 17.8 | 0.0536 | 0.0283 | 16.7 | 2.5 |
| 2019 | 30 | 4 | 625912 | 2 | 373843 | 7 | -22768 | 6.872 | 4.634 | 51 | 13.0 | 0.0473 | 0.0263 | 6.2 | 1.9 |
| 2019 | 30 | 5 | 506092 | 2 | 255402 | 7 | -10769 | 7.533 | 3.971 | 35 | 12.0 | 0.0453 | 0.0284 | 4.1 | 1.7 |
| 2019 | 30 | 6 | 421193 | 2 | 225456 | 7 | -18557 | 6.762 | 3.127 | 96 | 15.7 | 0.0546 | 0.0378 | 12.1 | 2.2 |
| 2019 | 30 | 7 | 363673 | 2 | 193717 | 7 | -14942 | 6.979 | 2.476 | 103 | 16.2 | 0.0481 | 0.0302 | 13.1 | 2.3 |
| 2019 | 30 | 8 | 432669 | 2 | 228125 | 7 | 12222 | 6.679 | 1.921 | 176 | 20.5 | 0.0486 | 0.0326 | 22.8 | 2.8 |
| 2019 | 30 | 9 | 526474 | 2 | 342438 | 7 | 43672 | 6.065 | 1.779 | 130 | 17.7 | 0.0455 | 0.0388 | 16.6 | 2.5 |
| 2019 | 30 | 10 | 954049 | 2 | 653576 | 7 | 48926 | 7.868 | 2.798 | 83 | 15.0 | 0.0973 | 0.0358 | 10.5 | 2.1 |
| 2019 | 30 | 11 | 951672 | 2 | 660333 | 7 | 33648 | 6.904 | 4.067 | 26 | 11.5 | 0.0972 | 0.0478 | 2.9 | 1.7 |
| 2019 | 30 | 12 | 963146 | 2 | 638716 | 7 | 29661 | 6.941 | 4.611 | 39 | 12.3 | 0.0830 | 0.0495 | 4.7 | 1.8 |
| 2020 | 31 | 1 | 944102 | 2 | 649373 | 9 | 33648 | 7.025 | 4.814 | 41 | 12.8 | 0.0870 | 0.0488 | 4.7 | 1.8 |
| 2020 | 31 | 2 | 1767958 | 2 | 1192155 | 9 | 73743 | 4.725 | 4.563 | 76 | 15.8 | 0.0919 | 0.0557 | 9.1 | 2.2 |
| 2020 | 31 | 3 | 1367473 | 2 | 812148 | 9 | 2981 | 5.120 | 4.053 | 100 | 17.9 | 0.0664 | 0.0578 | 12.1 | 2.5 |
| 2020 | 31 | 4 | 701215 | 2 | 400996 | 9 | -14905 | 6.477 | 3.687 | 41 | 12.7 | 0.0601 | 0.0222 | 4.7 | 1.8 |
| 2020 | 31 | 5 | 587604 | 2 | 300906 | 9 | -20867 | 7.124 | 3.372 | 28 | 11.7 | 0.0611 | 0.0295 | 3.1 | 1.7 |
| 2020 | 31 | 6 | 508823 | 2 | 249716 | 9 | -9427 | 6.670 | 3.080 | 74 | 15.7 | 0.0805 | 0.0759 | 8.9 | 2.2 |
| 2020 | 31 | 7 | 494290 | 2 | 246977 | 9 | -1714 | 6.192 | 2.621 | 80 | 16.2 | 0.1165 | 0.0272 | 9.6 | 2.3 |
| 2020 | 31 | 8 | 463982 | 2 | 239958 | 9 | -1751 | 5.887 | 2.400 | 134 | 20.9 | 0.0911 | 0.0280 | 16.5 | 2.9 |
| 2020 | 31 | 9 | 427740 | 2 | 222009 | 9 | -8347 | 6.358 | 2.121 | 99 | 17.9 | 0.0511 | 0.0302 | 12.1 | 2.5 |
| 2020 | 31 | 10 | 483024 | 2 | 249480 | 9 | 27463 | 7.059 | 2.297 | 65 | 14.8 | 0.0467 | 0.0321 | 7.7 | 2.1 |
| 2020 | 31 | 11 | 572292 | 2 | 296205 | 9 | 17215 | 6.410 | 2.535 | 22 | 11.1 | 0.0656 | 0.0394 | 2.3 | 1.6 |
| 2020 | 31 | 12 | 535572 | 2 | 347927 | 9 | 31413 | 7.229 | 3.079 | 32 | 12.0 | 0.0661 | 0.0397 | 3.6 | 1.7 |
| 2021 | 32 | 1 | 994408 | 2 | 654066 | 6 | 27947 | 8.434 | 4.224 | 41 | 12.2 | 0.0705 | 0.0439 | 5.2 | 1.7 |
| 2021 | 32 | 2 | 665684 | 2 | 370317 | 6 | 6297 | 7.474 | 4.685 | 76 | 15.0 | 0.0638 | 0.0448 | 10.0 | 2.1 |
| 2021 | 32 | 3 | 736993 | 2 | 467631 | 6 | 13340 | 7.033 | 4.409 | 101 | 17.1 | 0.0593 | 0.0297 | 13.5 | 2.3 |
| 2021 | 32 | 4 | 584913 | 2 | 310770 | 6 | -13974 | 7.294 | 4.600 | 41 | 12.1 | 0.0543 | 0.0158 | 5.1 | 1.7 |
| 2021 | 32 | 5 | 630061 | 2 | 382618 | 6 | 24109 | 5.952 | 4.228 | 28 | 11.1 | 0.0828 | 0.0178 | 3.4 | 1.6 |
| 2021 | 32 | 6 | 517628 | 2 | 271264 | 6 | -21650 | 6.624 | 3.713 | 75 | 14.9 | 0.0765 | 0.0296 | 9.8 | 2.0 |
| 2021 | 32 | 7 | 484386 | 2 | 295042 | 6 | -2236 | 5.688 | 3.008 | 81 | 15.4 | 0.0646 | 0.0456 | 10.6 | 2.1 |
| 2021 | 32 | 8 | 654433 | 2 | 390872 | 6 | 9092 | 5.439 | 2.198 | 136 | 19.9 | 0.0549 | 0.0295 | 18.3 | 2.7 |
| 2021 | 32 | 9 | 413749 | 2 | 314772 | 6 | 14011 | 6.225 | 2.170 | 101 | 17.0 | 0.0614 | 0.0273 | 13.4 | 2.3 |
| 2021 | 32 | 10 | 616283 | 2 | 463188 | 6 | 27127 | 7.115 | 2.709 | 65 | 14.1 | 0.0627 | 0.0384 | 8.5 | 1.9 |
| 2021 | 32 | 11 | 628364 | 2 | 387810 | 6 | 19041 | 7.515 | 3.359 | 22 | 10.5 | 0.0451 | 0.0432 | 2.4 | 1.5 |
| 2021 | 32 | 12 | 802914 | 2 | 563289 | 6 | 29400 | 7.903 | 4.069 | 32 | 11.4 | 0.0475 | 0.0414 | 3.9 | 1.6 |